\documentclass[prb,twocolumn,superscriptaddress,floatfix,longbibliography]{revtex4-2}

\usepackage{amsmath,amssymb,bm}
\usepackage{mathtools}
\usepackage{graphicx}
\usepackage{xcolor}
\usepackage{tikz}
\usetikzlibrary{arrows.meta}
\usepackage[colorlinks=true,linkcolor=blue,citecolor=blue,urlcolor=blue]{hyperref}

\newcommand{\Vd}{V_{\mathrm d}}
\newcommand{\Vs}{V_{\mathrm s}}
\newcommand{\Vu}{V_{\mathrm u}}
\newcommand{\gd}{g_{\mathrm d}}
\newcommand{\cv}{c}
\newcommand{\Jb}{\bar J}
\newcommand{\Tr}{T_1}
\newcommand{\Oq}{\Delta x}
\begin{document}

\title{Symmetry-resolved spectroscopy of truncation-induced perturbations\texorpdfstring{\\}{ }
  in loop tensor network renormalization of the six-vertex model}

\author{Akira Otsuki}
\email{akira-ots@g.ecc.u-tokyo.ac.jp}
\affiliation{Department of Applied Physics, University of Tokyo, Tokyo 113-8656, Japan}
\author{Yuta Yamamoto}
\email{4345823097@g.ecc.u-tokyo.ac.jp}
\affiliation{Department of Physics, University of Tokyo, Tokyo 113-0033, Japan}
\author{Kokin Cho}
\email{cho-kokin446@g.ecc.u-tokyo.ac.jp}
\affiliation{Department of Physics, University of Tokyo, Tokyo 113-0033, Japan}
\author{Rikuto Fushio}
\email{fushio-rikuto835@g.ecc.u-tokyo.ac.jp}
\affiliation{Department of Applied Physics, University of Tokyo, Tokyo 113-8656, Japan}
\author{Masahito Yamazaki}
\email{yamazaki@hep-th.phys.s.u-tokyo.ac.jp}
\thanks{Corresponding author.}
\affiliation{Department of Physics, University of Tokyo, Tokyo 113-0033, Japan}
\affiliation{Kavli Institute for Physics and Mathematics of the Universe (Kavli IPMU), UTIAS, University of Tokyo, Chiba 277-8583, Japan}
\affiliation{Center for Data-Driven Discovery (CD3), Kavli IPMU, UTIAS, University of Tokyo, Kashiwa, Chiba 277-8583, Japan}
\affiliation{Trans-Scale Quantum Science Institute, University of Tokyo, Tokyo 113-0033, Japan}
\affiliation{Center for Interdisciplinary Theoretical and Mathematical Sciences (iTHEMS), RIKEN, Saitama 351-0198, Japan}

\begin{abstract}
  Tensor network renormalization at finite bond dimension $\chi$ introduces truncation errors that are amplified by the renormalization-group (RG) flow.
  For loop tensor network renormalization (LoopTNR) on the $\mathsf{c}=1$ critical line of the six-vertex model, with Luttinger parameter $K$, we identify these errors operator by operator from symmetry-resolved transfer-matrix spectra and conformal perturbation theory.
  With $O(2)$ symmetry imposed, the dominant error is the dimerization operator, which is allowed because LoopTNR breaks one-site translation.
  With less symmetry it is the $U(1)$ current or a charge-two vertex operator.
  The growth of the dimerization per RG step and a parameter-free ratio of its amplitudes agree with conformal perturbation theory to within a few percent.
  The parameter $K$ drifts slowly, owing to a weak marginal error and to the running induced by the dimerization.
  A larger $\chi$ lowers the initial amplitude of the error for $0.75\le K\le1.5$ but changes neither the operator nor its growth rate.
\end{abstract}

\maketitle

\section{Introduction}
\label{sec:intro}

Tensor-network renormalization implements the renormalization group (RG) numerically in real space.
Starting from the tensor renormalization group (TRG)~\cite{Levin2007} and its higher-order variant~\cite{Xie2012}, it has developed into algorithms that remove short-range entanglement at each step, such as tensor network renormalization (TNR)~\cite{Evenbly2015}, loop tensor network renormalization (LoopTNR)~\cite{Yang2017}, and graph-independent local truncations~\cite{Hauru2018}.
At criticality these algorithms approximately realize a fixed-point tensor, from which the scaling dimensions of the underlying conformal field theory (CFT)~\cite{Gu2009,Cardy1984,Bloete1986,CardyNPB1986,Lyu2021}, operator product expansion coefficients~\cite{UedaYamazaki2023}, conformal defects~\cite{Hauru2016}, and RG flows~\cite{Bal2017} can be obtained.

At any finite bond dimension $\chi$, however, each truncation introduces a small error, which the RG flow amplifies.
Ueda and Oshikawa~\cite{Ueda2023} showed that the resulting deviations of the scaling dimensions are described by conformal perturbation theory, with the finite-$\chi$ error acting as a relevant perturbation whose growth can be read off from the transfer-matrix spectrum.
Tensor-network level spectroscopy has also resolved the Berezinskii--Kosterlitz--Thouless (BKT)~\cite{Berezinskii1971,Berezinskii1972,Kosterlitz1973} transition of the XY model~\cite{Ueda2021}.
For matrix product states, the length scale $\xi_\chi\propto\chi^{\kappa}$ set by finite entanglement~\cite{Tagliacozzo2008,Pollmann2009} has likewise been described as a relevant perturbation selected by the symmetries of the ansatz~\cite{Huang2024,Schneider2024}.

Critical lines with central charge $\mathsf{c}=1$ pose an additional difficulty.
The compact boson has an exactly marginal operator, so a numerical RG can not only leave the line of fixed points but also drift along it, changing the Luttinger parameter $K$.
Fixed-point tensors on such a line have been constructed analytically~\cite{Cheng2026}, and constructing and stabilizing critical fixed-point tensors is an active topic~\cite{Ebel2025,Gu2026b}.
In practice, the six-vertex benchmark of TNRKit.jl~\cite{TNRKit} stops LoopTNR at $\chi=40$ after 24 RG steps ``to avoid numerical errors arising from finite-bond-dimension effects.''
Which errors these are, and when they set in, has not been characterized.

In this paper we identify these errors.
We use the six-vertex model on its critical line as a lattice realization of the compact boson with tunable $K$, vary the symmetry imposed on the LoopTNR tensors, and ask in which operator direction a finite-$\chi$ run leaves the line of fixed points and how fast.
To identify the operator we use, in addition to the growth rate used in Ref.~\cite{Ueda2023}, a parameter-free ratio of amplitudes, a first-order level repulsion, and the phase of a spin-one coupling.
Table~\ref{tab:summary} summarizes the results.

With $O(2)$ imposed, the dominant error is the dimerization operator of dimension $K$.
It is allowed because LoopTNR does not preserve the original one-site translation, which becomes a fractional shift of the coarse lattice.
A shift along the critical line is not measurably amplified or damped, and the slow drift of $K$ that we observe comes from a weak marginal source and from the running induced by the growing dimerization.
With only $U(1)$ imposed, the $U(1)$ current appears in addition, with a coupling that rotates by $135^\circ$ per step, as expected for a spin-one operator.
With only charge conjugation $C$, the leading $U(1)$-breaking error is the charge-two vertex operator $V_{0,\pm2}$.
The bond dimension changes the initial amplitude of the error, which we call the seed, but neither the operator nor its growth rate.
A companion paper~\cite{Paper1b} will show how the seed fixes the correlation length that a finite-$\chi$ run eventually reaches, and how it can be measured, including its sign, and canceled.

\begin{table*}[tb]
  \caption{Dominant finite-$\chi$ perturbation for each imposed symmetry: operator and dimension, the observable used, the predicted change per RG step, and the measured value divided by the prediction ($\chi=40$ unless noted). For $C$ only, the more relevant $V_{0,\pm1}$ is permitted as well but enters the splitting only at second order and is visible in 6 of the 18 runs (Sec.~\ref{sec:conly}). For $U(1)$ only, the real part of the current coupling shifts the charged levels like a chemical potential, without opening a gap, and for $K\ge1.25$ this shift ends the window in which the charged levels can be read. For $K\le1$ the window is instead ended by dimerization (Sec.~\ref{sec:u1}).}
  \label{tab:summary}
  \begin{ruledtabular}
    \begin{tabular}{lllll}
      Symmetry & Operator ($x$)       & Observable                        & Prediction per step            & Measured                                                            \\
      \hline
      $O(2)$   & $\Vd$ ($K$)          & charged level, 2nd order          & $2^{2-K}$                      & $0.980$--$1.027$ ($0.75\le K\le1.75$)                               \\
               &                      & two-level amplitude ratio         & Eq.~\eqref{eq:ampratio}        & $0.998$--$1.005$ ($0.75\le K\le1.5$)                                \\
               &                      & charged level ($K>2$)             & no growth                      & none at $K=2.1$--$2.5$ (80 steps)                                   \\
      $U(1)$   & $J,\Jb$ ($1$)        & charge splitting, 1st order       & $\sqrt2$, rotation $135^\circ$ & rate$/\sqrt2=0.996$--$1.000$; angle $135.0^\circ$--$135.2^\circ$ \\
      $C$      & $V_{0,\pm2}$ ($1/K$) & $C$-even/odd splitting, 1st order & $2^{1-1/(2K)}$                 & $0.991$--$1.021$ (10 of 12 runs, $\chi=32$, $40$)                   \\
    \end{tabular}
  \end{ruledtabular}
\end{table*}

Section~\ref{sec:model} classifies the perturbations allowed by each imposed symmetry, and Sec.~\ref{sec:method} describes the LoopTNR setup, the transfer-matrix readout, and the relation between couplings and level shifts.
Sections~\ref{sec:o2} and \ref{sec:reduced} identify the dominant perturbation with $O(2)$ and with reduced symmetry, Sec.~\ref{sec:chi} discusses the bond dimension, and Sec.~\ref{sec:discussion} summarizes and discusses the results.

\section{Model and symmetry-allowed perturbations}
\label{sec:model}

\subsection{Six-vertex model and compact boson}

\begin{figure}[tb]
  \centering
  \begin{tikzpicture}[
      x=1cm,y=1cm,
      ice bond/.style={draw=black!45,line width=.55pt},
      ice arrow/.style={draw=black,line width=.7pt,
            -{Stealth[length=1.6mm,width=1.25mm]}}
    ]
    \node at (0,0.80) {$a=1$};
    \node at (1.45,0.80) {$b=1$};
    \node at (2.90,0.80) {$\cv$};
    \foreach \vx/\vy/\leftin/\rightin/\bottomin/\topin in {
        0/0/1/0/1/0,
        0/-1.38/0/1/0/1,
        1.45/0/1/0/0/1,
        1.45/-1.38/0/1/1/0,
        2.90/0/1/1/0/0,
        2.90/-1.38/0/0/1/1
      } {
        \begin{scope}[shift={(\vx,\vy)}]
          \draw[ice bond] (-.58,0) -- (.58,0);
          \draw[ice bond] (0,-.58) -- (0,.58);
          \foreach \angle/\incoming in {
              180/\leftin,0/\rightin,270/\bottomin,90/\topin
            } {
              \ifnum\incoming=1
                \draw[ice arrow] (\angle:.50) -- (\angle:.18);
              \else
                \draw[ice arrow] (\angle:.18) -- (\angle:.50);
              \fi
            }
          \fill (0,0) circle[radius=1.55pt];
        \end{scope}
      }
  \end{tikzpicture}
  \caption{The six vertex configurations allowed by the ice rule and their Boltzmann weights. The two configurations in each column are related by reversal of all arrows. We set $a=b=1$ and vary $\cv$.}
  \label{fig:vertices}
\end{figure}
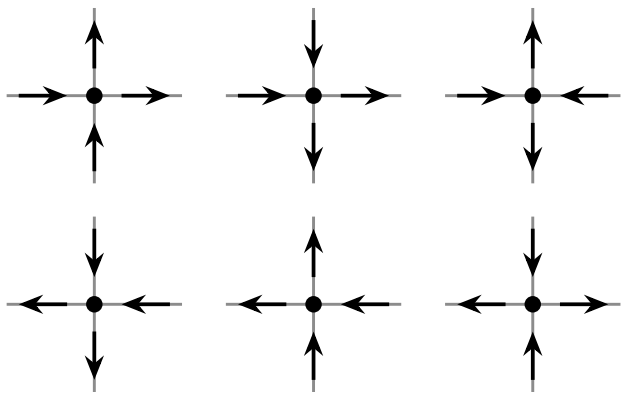

The six-vertex model assigns an arrow to each bond of the square lattice, subject to the ice rule that two arrows point into and two out of each vertex~\cite{Lieb1967,Baxter1982}.
We give equal weights to configurations related by reversal of all arrows and set $(a,b,\cv)=(1,1,\cv)$ (Fig.~\ref{fig:vertices}), which defines the F model~\cite{Rys1963,Lieb1967F}.
Antiferroelectric order sets in only for $\cv>2$, and the whole range $0<\cv<2$ forms a single critical phase, which we parametrize by
\begin{equation}
  \Delta_{6\mathrm v}\coloneqq1-\frac{\cv^2}{2},\qquad
  K\coloneqq\frac{\pi}{2\arccos\Delta_{6\mathrm v}},
  \label{eq:K}
\end{equation}
or equivalently $\cv=2\sin[\pi/(4K)]$, so that $K>1/2$ in this convention.
The free-fermion point is $K=1$ ($\cv=\sqrt2$), and $K\to1/2$ ($\cv\to2$) is the BKT point of the F model.
In the continuum the model becomes a compact boson with $\mathsf{c}=1$~\cite{Kadanoff1979,Nienhuis1984,DiFrancesco1987,Alcaraz1988}, whose radius is an exactly marginal modulus~\cite{Ginsparg1988,Dijkgraaf1988}.

Primary states carry two integer charges, the $U(1)$ charge $m$ and the dual charge $e$, and in terms of
\begin{equation}
  p\coloneqq e\sqrt K+\frac{m}{2\sqrt K},\qquad
  \bar p\coloneqq e\sqrt K-\frac{m}{2\sqrt K},
  \label{eq:charges}
\end{equation}
the vertex operator $V_{e,m}$ has scaling dimension and conformal spin
\begin{equation}
  x_{e,m}=\frac{p^2+\bar p^2}{2}=e^2K+\frac{m^2}{4K},\qquad s_{e,m}=\frac{p^2-\bar p^2}{2}=em .
  \label{eq:spectrum}
\end{equation}
Descendants, with $x=x_{e,m}+N+\bar N$ and $s=em+N-\bar N$ ($N,\bar N=0,1,2,\dots$), are generated by the modes $J_n$, $\Jb_n$ ($n<0$) of the $U(1)$ currents $J(z)=\sum_nJ_nz^{-n-1}$ and $\Jb(\bar z)=\sum_n\Jb_n\bar z^{-n-1}$.
The modes satisfy $[J_n,J_{n'}]=n\,\delta_{n+n',0}$, and the primary state $|e,m\rangle$ of $V_{e,m}$ has $J_0|e,m\rangle=p\,|e,m\rangle$ (likewise for $\Jb_n$ and $\bar p$).
We write the boson as $\phi=\phi_L(z)+\phi_R(\bar z)$, with $\langle\phi_L(z)\phi_L(w)\rangle=-\ln(z-w)$ and likewise for $\phi_R$.
Then $J=i\partial\phi$, $\Jb=i\bar\partial\phi$, and $V_{e,m}={:}e^{i(p\phi_L+\bar p\phi_R)}{:}={:}e^{ie\sqrt K\phi+im\tilde\phi/(2\sqrt K)}{:}$, where $\tilde\phi=\phi_L-\phi_R$ is the dual field.
In this normalization $e$ and $m$ are the momentum and winding numbers of $\phi$, whose compactification radius is $R=1/\sqrt K$.
The field $\phi$ is proportional to the coarse-grained height of the six-vertex model (Appendix~\ref{app:lattice-sym}).
For the finite-size spectrum we map the plane to a cylinder of circumference $L$ by $z=e^{2\pi w/L}$ with $w=t_E+iy$, where the imaginary time $t_E$ runs along the cylinder and $y$ around it.
The vertex operators are normalized as $\langle V_{e,m}(z,\bar z)V_{-e,-m}(0)\rangle=z^{-p^2}\bar z^{-\bar p^2}$.

\subsection{Symmetries and allowed operators}
\label{sec:symmetry}

Charge conjugation $C$ maps $(e,m)\mapsto(-e,-m)$ and reverses the sign of the $U(1)$ currents $J,\Jb$.
One-site translation $\Tr$ acts as $V_{e,m}\mapsto(-1)^eV_{e,m}$, so at long distances it is an internal $\mathbb Z_2$ (Appendix~\ref{app:lattice-sym}).
The fourfold rotation $C_4$, by $90^\circ$ about the center of a face, multiplies an operator of spin $s$ by $i^s$.
A perturbation must therefore have $m=0$ to preserve $U(1)$ and be $C$-even to preserve $C$, while $\Tr$ forbids odd $e$ and $C_4$ forbids $s\not\equiv0\pmod4$.
Point-group operations that exchange the two sublattices of faces, such as a rotation by $90^\circ$ about a vertex or a reflection about a line through a column of vertices, also forbid the dimerization field $\Vd$ defined below.
In the scheme used here, however, no point-group symmetry is imposed on the coarse-grained tensors (Appendix~\ref{app:lattice-sym}).

These rules are those of the bosonization of the XXZ chain and of the Coulomb gas~\cite{Haldane1980,Haldane1981,Haldane1982,Kadanoff1979,Nienhuis1984,Alcaraz1988,Giamarchi2004}, and level spectroscopy uses the same quantum numbers~\cite{Okamoto1992,Nomura1994,Nomura1995}.
Table~\ref{tab:operators} collects them, and Appendix~\ref{app:lattice-sym} derives them in the arrow basis of the tensors, in which each bond index labels the direction of the arrow on that bond (the $S^z$ basis of the XXZ chain).
Of these symmetries, $U(1)$ and $C$ can be imposed on LoopTNR tensors, whereas no practical LoopTNR preserving $\Tr$ is known.
The perturbations that matter most below are the dimerization, staggered, and umklapp fields
\begin{equation}
  \begin{gathered}
    \Vd\coloneqq\frac{V_{1,0}+V_{-1,0}}{\sqrt2},\qquad
    \Vs\coloneqq\frac{V_{1,0}-V_{-1,0}}{i\sqrt2},\\
    \Vu\coloneqq\frac{V_{2,0}+V_{-2,0}}{\sqrt2},
  \end{gathered}
  \label{eq:Vd}
\end{equation}
of dimensions $K$, $K$, and $4K$, respectively.
In the XXZ chain, $\Vd$ is the bond alternation (Appendix~\ref{app:coefficients}).

\begin{table*}[tb]
  \caption{Classification of low-dimension perturbations by the symmetries of the model. $\circ$: allowed; $\times$: forbidden; ``even'': only the $C$-even combination is allowed. $C_4$ is the rotation by $90^\circ$ about the center of a face (a rotation about a vertex gives an additional factor $(-1)^e$). $x_j<2$ is relevant and $x_j=2$ marginal. $T$ and $\bar T$ are the holomorphic and antiholomorphic stress tensors, $(TT)$ denotes the conformal normal-ordered product (the regular part of the $T(z)T(w)$ OPE), $\Lambda\coloneqq(TT)-\tfrac{3}{10}\partial^2T$, $W\coloneqq{:}J^4{:}$, and $J^3\Jb\coloneqq{:}J^3{:}\Jb$, with free-field normal ordering; bars denote antiholomorphic counterparts. For $V_{0,\pm1}$, $V_{0,\pm2}$, and $\Vu$, $g$ and $g'$ are the couplings of the $C$-even and $C$-odd combinations, defined as for $\Vd$ and $\Vs$ in Eq.~\eqref{eq:Vd}; all couplings are normalized as in Eq.~\eqref{eq:coupling}. The table is compiled from the transformation rules of Sec.~\ref{sec:symmetry}, derived in Appendix~\ref{app:lattice-sym}. Descendants of $V_{e,m}$ with $(e,m)\neq(0,0)$ are not listed.}
  \label{tab:operators}
  \begin{ruledtabular}
    \begin{tabular}{llcccccc}
      Operator                                   & Coupling                       & $x_j$      & $s$      & $U(1)$   & $C$      & $\Tr$    & $C_4$    \\
      \hline
      $V_{0,\pm1}$                               & $g_{0,1},g'_{0,1}$             & $1/(4K)$   & $0$      & $\times$ & even     & $\circ$  & $\circ$  \\
      $V_{0,\pm2}$                               & $g_{0,2},g'_{0,2}$             & $1/K$      & $0$      & $\times$ & even     & $\circ$  & $\circ$  \\
      other $V_{e,m}$, $m\neq0$                  & ---                            & $x_{e,m}$  & $em$     & $\times$ & even     & $e$ even & $em\in4\mathbb Z$ \\
      $J,\Jb$                                    & $g_J,g_{\Jb}$                  & $1$        & $\pm1$   & $\circ$  & $\times$ & $\circ$  & $\times$ \\
      $\Vd$                                      & $\gd$                          & $K$        & $0$      & $\circ$  & $\circ$  & $\times$ & $\circ$  \\
      $\Vs$                                      & $g_{\mathrm s}$                & $K$        & $0$      & $\circ$  & $\times$ & $\times$ & $\circ$  \\
      $T,\bar T$                                 & $g_T,g_{\bar T}$               & $2$        & $\pm2$   & $\circ$  & $\circ$  & $\circ$  & $\times$ \\
      $J\Jb$                                     & $g_{J\Jb}$                     & $2$        & $0$      & $\circ$  & $\circ$  & $\circ$  & $\circ$  \\
      $\Vu$                                      & $g_{\mathrm u},g'_{\mathrm u}$ & $4K$       & $0$      & $\circ$  & even     & $\circ$  & $\circ$  \\
      $T\bar T,\ \Lambda+\bar\Lambda,\ W+\bar W$ & $g_{T\bar T},g_\Lambda,g_W$    & $4$        & $0,\pm4$ & $\circ$  & $\circ$  & $\circ$  & $\circ$  \\
      $J^3\Jb,\ J\Jb^3$                          & ---                            & $4$        & $\pm2$   & $\circ$  & $\circ$  & $\circ$  & $\times$ \\
    \end{tabular}
  \end{ruledtabular}
\end{table*}

If both $O(2)=U(1)\rtimes\mathbb Z_2^C$ and $\Tr$ are preserved, no relevant perturbation is allowed for $K>1/2$.
The leading corrections are then the exactly marginal $J\Jb$, which shifts $K$, the marginal $T,\bar T$ if $C_4$ is not imposed, the umklapp $\Vu$, and dimension-four operators.
In LoopTNR with $O(2)$ imposed, however, $\Tr$ is lost, because the coarse-grained network has a two-tensor unit cell and an internal symmetry of the tensors does not enforce invariance under the original one-site translation.
This allows $\Vd$, which is relevant for $K<2$.
We expect that neither imposing equal tensors in the unit cell nor imposing $C_4$ restores $\Tr$, which after $n$ steps has no local representation on the coarse network [Fig.~\ref{fig:schematic}(a) and Appendix~\ref{app:lattice-sym}].
We have not run such schemes.
Relaxing the symmetry admits more operators.
With only $U(1)$ imposed, $\Vs$ and the currents $J,\Jb$ are allowed, and the currents, of dimension one, are relevant for all $K$, although their uniform part acts only on the zero modes (Sec.~\ref{sec:u1}).
The $C$-odd combination of $V_{\pm2,0}$ (coupling $g'_{\mathrm u}$ in Table~\ref{tab:operators}) is allowed as well but is irrelevant for $K>1/2$.
With only $C$ imposed, $\Vs$ and the currents are forbidden, but the $C$-even combinations of all $V_{e,m}$ with $m\neq0$ are allowed.
Of these, only $V_{0,\pm1}$ and $V_{0,\pm2}$, of dimensions $1/(4K)$ and $1/K$, enter the analysis below.

\section{Method}
\label{sec:method}

\subsection{LoopTNR and transfer-matrix spectroscopy}
\label{sec:terms}

\begin{figure*}[tb]
  \centering
  \includegraphics[width=\textwidth]{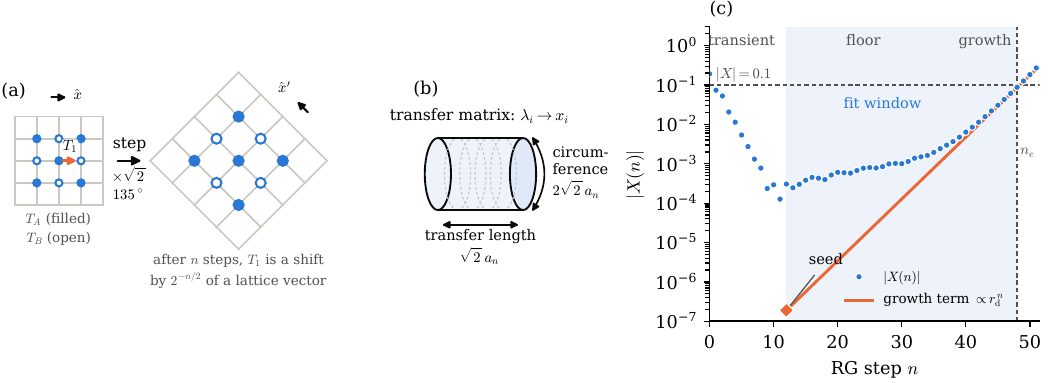}
  \caption{(a) One RG step of LoopTNR: the network with a two-tensor unit cell ($T_A$, $T_B$) is coarse-grained to a lattice rescaled by $\sqrt2$ and rotated by $135^\circ$, as indicated by the direction of the same tensor leg before ($\hat x$) and after ($\hat x'$) the step. The original one-site translation $\Tr$ becomes a shift by a fraction $2^{-n/2}$ of a coarse lattice vector. (b) Readout tube: a cylinder of circumference $2\sqrt2\,a_n$ and transfer length $\sqrt2\,a_n$ built from the step-$n$ tensors. (c) Relative deviation $|X(n)|$ of the lowest charged level ($O(2)$, $K=1.5$, $\chi=40$): short-distance transient, floor, growth of the dimerization term, and breakdown at $|X|=0.1$ ($n_e$ is the last step before it). The fit window and the extrapolation of the growth term back to $n=12$, which defines the seed, are indicated.}
  \label{fig:schematic}
\end{figure*}

We use the LoopTNR algorithm~\cite{Yang2017} as implemented in TNRKit.jl~\cite{TNRKit}, with the tensors built in the TensorKit.jl symmetric-tensor framework~\cite{TensorKit,Singh2010,Singh2011,Singh2012} so that either $U(1)$ or $O(2)$ is imposed exactly.
Section~\ref{sec:conly} uses a variant of this setup in which only $C$ is imposed.
No spatial symmetry is imposed.
In particular, the two tensors $T_A$ and $T_B$ of the unit cell [Fig.~\ref{fig:schematic}(a)] are optimized independently, and neither $C_4$ rotation nor reflections are enforced.
Each RG step rescales lengths by $\sqrt2$, giving the lattice spacing $a_n=2^{n/2}a_0$ after $n$ steps.
In this implementation it also rotates the lattice by $135^\circ$ counterclockwise~\cite{TNRKitCode} [Fig.~\ref{fig:schematic}(a)].
The angle matters because $C_4$ is not imposed (Appendix~\ref{app:numerics}), and it can be seen directly in the phase of the current coupling in Sec.~\ref{sec:u1}.
We use $\chi=40$ for the main analysis and $\chi=16$, $24$, $32$, $48$, $64$, and $96$ to study the $\chi$ dependence.
Unless stated otherwise, the data come from the main runs, which use a single released code version and identical parameters for all $K$ and $\chi$; all data sets are listed in Appendix~\ref{app:numerics}.

At each step $n$ we contract the coarse-grained tensors into a cylinder of circumference $2\sqrt2\,a_n$ and transfer length $\sqrt2\,a_n$ with zero twist, which we call the \emph{standard tube} [Fig.~\ref{fig:schematic}(b)].
This is the default readout of TNRKit~\cite{TNRKitCode}.
Only the circumference sets the finite-size spectrum, and the transfer length merely fixes the power of the transfer matrix.
We then compute the leading eigenvalues $\lambda_i$ of its transfer matrix in each symmetry sector.
The scaling dimensions are read off as~\cite{Cardy1984,Gu2009}
\begin{equation}
  x_i(n)\coloneqq-\frac{1}{2\pi\operatorname{Im}\tau_n}\ln\left|\frac{\lambda_i}{\lambda_0}\right| ,
  \label{eq:readout}
\end{equation}
where $\tau_n$ is the modular parameter of the tube, equal to $[\textrm{(twist)}+i\times\textrm{(transfer length)}]/\textrm{(circumference)}$ for an isotropic tensor.
We fix it to this value, $\tau_n=i/2$.
A residual anisotropy of the coarse-grained tensor then enters as a common relative error $\eta$ of all levels (Sec.~\ref{sec:cpt}).
The central charge $\mathsf{c}$ is obtained from the leading eigenvalue $\lambda_0$ (Appendix~\ref{app:numerics}).
The eigenvalues $\lambda_i$ are in general complex.
Equation~\eqref{eq:readout} uses their moduli, while Sec.~\ref{sec:u1-phase} uses their phases.
As a check of the readout we also use a \emph{wide tube}, which coarse-grains the step-$n$ tensors once more and has $\sqrt5$ times the circumference and a small twist, $\tau=0.1+0.2i$ (Appendix~\ref{app:numerics}).
Levels are assigned to CFT states by their $U(1)$ charge, by their $C$ parity when $O(2)$ is imposed, and by the unperturbed ordering and degeneracies.
Because the untwisted tube does not resolve the conformal spin, spin-$\pm s$ partners are identified as nearly degenerate pairs, and we restrict the analysis to ranges of RG steps in which this assignment is unambiguous.

Figure~\ref{fig:schematic}(c) shows a typical history of the relative deviation of a level from its CFT value, here $X(n)$ of Eq.~\eqref{eq:X} for the lowest charged level.
Short-distance corrections die out during the first $\sim10$--$20$ steps.
The deviation then sits at a \emph{floor} of order $10^{-4}$, set by the readout and the slow drift of $K$, until the response to a relevant perturbation grows out of it.
Once $|X|$ exceeds $0.1$ the levels leave the CFT spectrum.
We call this the \emph{breakdown} and denote by $n_e$ the last step before it.
The steps up to $n_e$ form the \emph{weak-coupling window}, and the range of steps used in a given fit is its fit window.
We write $r_{\rm d}$ for the growth factor per step of the dimerization contribution to the levels and $r_J$ for that of the current.

\subsection{Couplings and level shifts}
\label{sec:cpt}

We describe the finite-size spectrum by conformal perturbation theory~\cite{Reinicke1987,Ueda2023} for the Hamiltonian on a circle of circumference $L$ that generates the transfer matrix along the tube,
\begin{equation}
  H=H_{\mathrm{CFT}}+\sum_j\tilde g_j\int_0^{L}dy\,\Phi_j(y),
\end{equation}
and define the dimensionless coupling of an operator of dimension $x_j$ as
\begin{equation}
  g_j(L)\coloneqq\tilde g_j\left(\frac{L}{2\pi}\right)^{2-x_j},\qquad L\coloneqq2\sqrt2\,a_n,
  \label{eq:coupling}
\end{equation}
the circumference of the standard tube at step $n$ in units of $a_0$.
In the weak-coupling regime $g_j(L)\propto L^{2-x_j}$, so that a contribution of order $k$ in $g_j$ to a level grows by
\begin{equation}
  r_j=2^{k(2-x_j)/2}
  \label{eq:rate}
\end{equation}
per RG step.
Since each step also rotates the lattice by $135^\circ$, the coupling of an operator with spin $s$ picks up a phase $e^{3\pi is/4}$ per step.
For a real tensor network the Euclidean action is real, and this relates the couplings of spin-$\pm s$ partners.
In particular, $g_{\bar T}=g_T^*$ and $g_{\Jb}=-g_J^*$, the latter because our convention $J=i\partial\phi$, $\Jb=i\bar\partial\phi$ implies $J^*=-\Jb$ for real $\phi$.
The resulting growth factors and phases are listed in Table~\ref{tab:rates}.

\begin{table}[tb]
  \caption{Predicted change per RG step of the contribution of each perturbation to a scaling dimension (Sec.~\ref{sec:cpt}). $k$ is the lowest order in the coupling at which the perturbation shifts the level; the factor per step is $2^{k(2-x_j)/2}$ [Eq.~\eqref{eq:rate}]. The phase per step is the rotation $s\times135^\circ$ (modulo $360^\circ$) of the coupling of an operator of spin $s$, due to the $135^\circ$ rotation of the lattice at each step; the upper (lower) sign refers to the holomorphic (antiholomorphic) member of a spin-$\pm s$ pair. A first-order contribution therefore varies as $r_j^n\cos(n\varphi+\varphi_0)$, where $r_j$ is the factor and $\varphi$ the phase per step: $180^\circ$ is a sign alternation, i.e., a factor $-\tfrac12$ per step [Eq.~\eqref{eq:fit}], and $\pm270^\circ$ gives period four. Second-order contributions are proportional to $|g_j|^2$ and do not rotate. The row ``$\Vu$ (degenerate levels)'' applies to levels connected by $\Vu$ to a degenerate partner, e.g., the $C$-even and $C$-odd combinations of $(\pm1,0)$.}
  \label{tab:rates}
  \begin{ruledtabular}
    \begin{tabular}{lccc}
      Source                    & $k$ & factor per step & phase per step \\
      \hline
      $\Vd$, $\Vs$              & 2   & $2^{2-K}$       & $0$            \\
      $J,\Jb$                   & 1   & $\sqrt2$        & $\pm135^\circ$ \\
      $\Vu$                     & 2   & $2^{2-4K}$      & $0$            \\
      $\Vu$ (degenerate levels) & 1   & $2^{1-2K}$      & $0$            \\
      $T\bar T$                 & 1   & $1/2$           & $0$            \\
      $\Lambda,W$ (spin $\pm4$) & 1   & $1/2$           & $180^\circ$    \\
      $J^3\Jb$ (spin $\pm2$)    & 1   & $1/2$           & $\pm270^\circ$ \\
      $T,\bar T$                & 1   & $1$             & $\pm270^\circ$ \\
      $J\Jb$ ($\delta K$)       & 1   & $1$             & $0$            \\
    \end{tabular}
  \end{ruledtabular}
\end{table}

For spin-zero levels, or spin-averaged pairs, of an $O(2)$-symmetric system the deviation from the CFT value $x_i^{\rm CFT}(K)$ at the input $K$ is, to leading order,
\begin{equation}
  \delta x_i=\eta\,x_i^{\rm CFT}+\partial_Kx_i^{\rm CFT}\,\delta K+\alpha_i(K)\,\gd^2+\sum_jM_{ij}g_j+\cdots .
  \label{eq:deltax}
\end{equation}
The first term is a common relative error $\eta$ of the level scale, arising, e.g., from the value of $\tau_n$.
It also absorbs the first-order effect of $T,\bar T$, which are allowed because $C_4$ is not imposed and which enter $\eta$ with period four in $n$.
The second term is the shift $\delta K$ along the critical line generated by $J\Jb$, with $\delta K=2\pi K g_{J\Jb}$ to first order.
The third term is the second-order shift by $\Vd$, with coefficient $\alpha_i$, and the last collects the first-order shifts by $\Vu$ and the dimension-four operators, with couplings $g_j$ and coefficients $M_{ij}$.
The dots include the second-order shift by $\Vu$, which gives $A_{\rm u}$ in Eq.~\eqref{eq:fit}.
Because $\Vd$ changes $e$ by $\pm1$, it has no diagonal matrix elements between nondegenerate states and enters at second order, except when it connects degenerate states (Sec.~\ref{sec:k1}).
The coefficients $\alpha_i(K)$ follow in closed form from the Coulomb-gas integrals (Appendix~\ref{app:coefficients}).
We extrapolate the growth term of the charged level [Eq.~\eqref{eq:fit}, Sec.~\ref{sec:o2-terms}] back to $n=12$ with the fitted growth factor and divide it by $-4K\alpha_{0,1}$.
This gives the dimerization coupling $\gd^2(12)$, which we call the \emph{seed}.
The companion paper~\cite{Paper1b} will show that it measures the error injected during the first steps after the first truncation.
Its value depends on the fit protocol by a $K$-dependent factor, but its dependence on $\chi$ does not (Appendix~\ref{app:numerics}).

When only $U(1)$ is preserved, the currents contribute already at first order,
\begin{equation}
  \delta x_i\supset2\pi\bigl(g_Jp_i+g_{\Jb}\bar p_i\bigr),
  \label{eq:current-shift}
\end{equation}
since $J(z)V_{e,m}(0)\sim pV_{e,m}(0)/z$ and $\Jb(\bar z)V_{e,m}(0)\sim\bar pV_{e,m}(0)/\bar z$, i.e., their zero modes are $p$ and $\bar p$ (Sec.~\ref{sec:model}).
For the primary states $(e,m)=(0,\pm q)$, Eq.~\eqref{eq:current-shift} gives $\delta x_{0,\pm q}=\pm(\pi q/\sqrt K)\,g_-$ with $g_-\coloneqq g_J-g_{\Jb}$.
This splits the levels of opposite charge linearly in $q$ and leaves their average unchanged at this order.
With $g_{\Jb}=-g_J^*$ the combination $g_-=2\operatorname{Re}g_J$ is real, and the splitting appears in $|\lambda_i|$.
The orthogonal combination $g_J+g_{\Jb}=2i\operatorname{Im}g_J$ couples to $J+\Jb\propto\partial_{t_E}\phi$ and shifts the states $(\pm e,0)$ by the imaginary amounts $\pm4\pi i\sqrt K e\operatorname{Im}g_J$.
These shifts show up in the phases of the transfer-matrix eigenvalues and do not enter Eq.~\eqref{eq:readout}.
As $g_J$ rotates by $135^\circ$ per step, the two parts alternate in quadrature.
The same assignment follows from the reality of the transfer matrix alone (Appendix~\ref{app:u1checks}).

Finally, in the weak-coupling regime the couplings $\gd$ and $K$ obey the RG equations~\cite{Kosterlitz1974,Giamarchi2004}
\begin{equation}
  \frac{d\gd}{d\ell}=(2-K)\gd+\cdots,\qquad
  \frac{dK}{d\ell}=-2\pi^2K^2\gd^2+\cdots,
  \label{eq:flow}
\end{equation}
with $\ell=\ln L$, in the normalization of Eq.~\eqref{eq:coupling}.
The second equation shows that a growing dimerization also drives $K$ downward.
The coefficients $\alpha_i(K)$ reflect this through a pole $\alpha_i\simeq-\pi^2K^2\partial_Kx_i/(2-K)$ as $K\to2$.
The residue of this pole, which two independently computed levels reproduce, fixes the coefficient $2\pi^2K^2$ in our normalization (Appendix~\ref{app:coefficients}).

Truncation errors added at every RG step enter these equations as sources,
\begin{equation}
  \frac{d\gd}{d\ell}=(2-K)\gd+\sigma_{\mathrm d}(\ell),\qquad
  \frac{dK}{d\ell}=-2\pi^2K^2\gd^2+\sigma_K(\ell),
  \label{eq:sourceflow}
\end{equation}
where $\sigma_{\mathrm d}$ and $\sigma_K$ are effective sources from the truncation.
Growth at a given rate can then come either from a seed injected in the first steps and amplified afterward or from sources acting at every step.
The companion paper~\cite{Paper1b} will separate the two by branching the runs at intermediate steps.

\section{\texorpdfstring{$O(2)$}{O(2)}-symmetric LoopTNR: dimerization}
\label{sec:o2}

With $O(2)$ imposed, the deviation of the levels from the CFT is dominated by a single growing term.
We show that it has the growth rate of a second-order dimerization (Sec.~\ref{sec:o2-rate}), the parameter-free amplitude ratio of $\Vd$ in two levels (Sec.~\ref{sec:o2-amp}), and, at $K=1$, the first-order level repulsion that $\Vd$ predicts (Sec.~\ref{sec:k1}).
Sections~\ref{sec:bkt} and \ref{sec:Kn} then treat the BKT point and the slow motion along the critical line.

\subsection{Charged level: decomposition of the deviation}
\label{sec:o2-terms}

We begin with the lowest level in the charge sector $|m|=1$, whose CFT dimension is $x_{0,1}=1/(4K)$, and follow its relative deviation
\begin{equation}
  X(n)\coloneqq1-4Kx_{0,1}(n).
  \label{eq:X}
\end{equation}
Guided by Eq.~\eqref{eq:deltax} and Table~\ref{tab:rates}, we model the signed $X(n)$ as
\begin{equation}
  \begin{aligned}
    X(n)\simeq{} & \beta_0+\beta_1n+A\,r_{\rm d}^{\,n-n_e}+A_{\rm u}(2^{2-4K})^{n-n_s} \\
            & +A_4(-\tfrac12)^{n-n_s}+A_0(\tfrac12)^{n-n_s}.
  \end{aligned}
  \label{eq:fit}
\end{equation}
Here the linear background absorbs $\eta$ and the slow drift of $K$, and $A$ is the dimerization term with a free rate $r_{\rm d}$.
The period-four part of $\eta$ from $T,\bar T$ has an amplitude of only $(3$--$5)\times10^{-5}$ (Appendix~\ref{app:k175}) and is not fitted.
The remaining terms are the second-order umklapp term $A_{\rm u}$ and the first-order responses $A_4$ and $A_0$ to the dimension-four operators of spin $\pm4$ and $0$.
Two further terms allowed without $C_4$ are not fitted: the first-order response to the spin-$\pm2$ operators $J^3\Jb$ and $J\Jb^3$, which decays by $1/2$ per step with period four, and, for $K<1$, the second-order response to the spin-$\pm2$ level-two descendants of $\Vd$, which decays by $2^{-K}$ per step.
Adding either term does not improve the fit significantly.
Because all other rates are fixed by theory, the amplitudes for each trial $r_{\rm d}$ follow from linear least squares.
The fit runs from $n_s=6$ to the end $n_e$ of the weak-coupling window.

\begin{figure*}[tb]
  \centering
  \includegraphics[width=\textwidth]{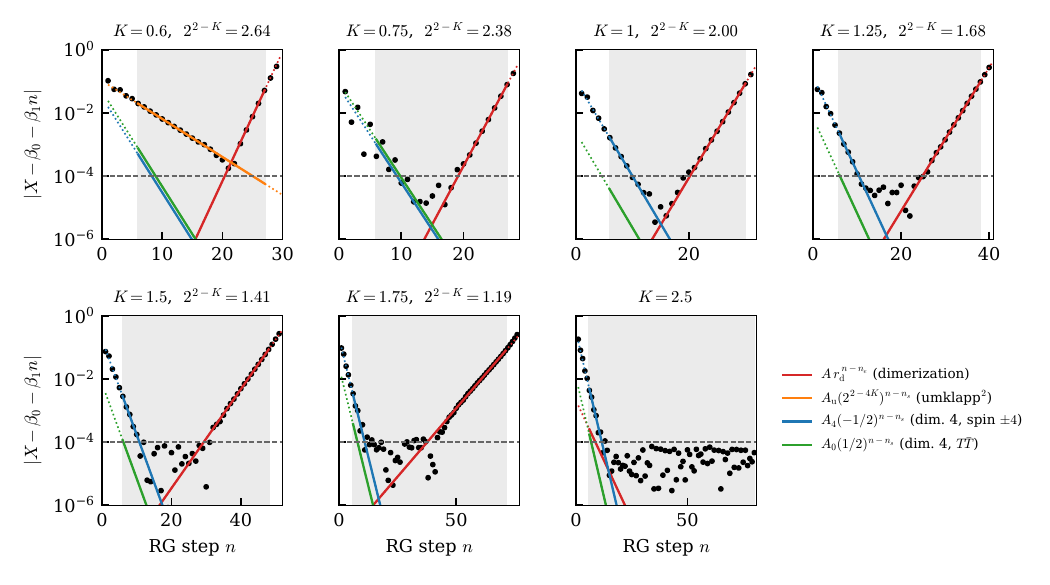}
  \caption{$O(2)$-symmetric LoopTNR at $\chi=40$: deviation of the lowest charged level, $X(n)=1-4Kx_{0,1}(n)$, after subtraction of the fitted linear background (dots), and the individual terms of Eq.~\eqref{eq:fit} with $n_s=6$ (lines; solid inside the fit window, shaded, and dotted outside). The dashed line marks $10^{-4}$, the typical level of the floor. For $K=2.5$ all rates are fixed to their theoretical values. $K=0.6$ is from an additional run with the same code and settings (Appendix~\ref{app:numerics}).}
  \label{fig:terms}
\end{figure*}

Figure~\ref{fig:terms} shows the result for seven values of $K$ at $\chi=40$.
For $0.75\le K\le1.75$ the deviation passes through the stages sketched in Fig.~\ref{fig:schematic}(c).
It settles on a floor, which lasts longer for larger $K$, until a growing term takes over and drives $|X|$ to $0.1$ after $27$ to $71$ steps.
At $K=2.5$, where $\Vd$ is irrelevant, nothing grows out of the floor within 80 steps.
Closer to the BKT point, at $K=0.6$, the umklapp term decays slowly, by $0.76$ per step, and dominates up to $n\approx22$.
The dimerization then takes over without an intermediate floor.
At $K=0.75$ the umklapp rate $2^{2-4K}=1/2$ coincides with that of $A_0$, and the two terms cannot be separated.

\subsection{Growth rate}
\label{sec:o2-rate}

\begin{figure}[tb]
  \centering
  \includegraphics[width=\columnwidth]{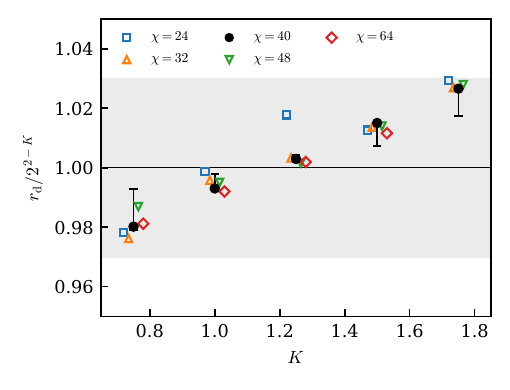}
  \caption{Growth factor per RG step $r_{\rm d}$ of the deviation of the lowest charged level, divided by the second-order dimerization prediction $2^{2-K}$, for $O(2)$ symmetry and $\chi=24$--$64$ (shifted horizontally for visibility). Error bars ($\chi=40$): range over fit windows with start $n_s=10,12,14$ and cut $|X|\le0.05,0.10$. The shaded band indicates $\pm3\%$. The rate fixes the dimension and the order of the perturbation, but by itself it does not identify the operator (see Fig.~\ref{fig:amp}).}
  \label{fig:rate}
\end{figure}

The growing term has the rate of a second-order dimerization, $2^{2-K}$ [Eq.~\eqref{eq:rate} with $x=K$ and $k=2$], to within $3\%$ for $0.75\le K\le1.75$ (Fig.~\ref{fig:rate} and Table~\ref{tab:o2}), and to within $1\%$ after extrapolation to weak coupling (median over $\chi\ge32$; each extrapolated run with $\chi\ge48$).
For $K=2.1$--$2.5$, where $\Vd$ is irrelevant, no growth appears within 80 steps (Appendix~\ref{app:numerics}).

To measure $r_{\rm d}$ we refit from $n_s=12$, where the dimension-four terms have fallen below the floor, keeping the background, the growth term, and, for $K<3/4$, the umklapp term.
Adding the $J^3\Jb$ term and, for $K<1$, the descendant term of Sec.~\ref{sec:o2-terms} to this refit changes $r_{\rm d}$ and the amplitude ratio of Sec.~\ref{sec:o2-amp} by less than $0.7\%$.
At $K=3/4$ the umklapp term has the rate of $A_0$ and, like $A_0$, lies below the floor from $n=12$ on (Fig.~\ref{fig:terms}).
At $\chi=40$ the ratio $r_{\rm d}/2^{2-K}$ increases from $0.980$ at $K=0.75$ to $1.027$ at $K=1.75$, and the values for $\chi=24$--$64$ lie within $1\%$ of these, except at $K=1.25$ and $\chi=24$ ($1.018$) and at $K=1.75$ and $\chi=64$, where the window exceeds the 80 steps of the run.
We quote as the uncertainty the spread over fit windows.
For the growth factor it is larger than the spread from a block bootstrap of the residuals in all cases (Appendix~\ref{app:numerics}).

The remaining deviations are systematic.
The rate is low for $K\le1$, by up to $2\%$ ($3\%$ at $K=0.6$ and $\chi=40$), and high for $K\ge1.5$, by up to $3\%$.
Both decrease as the window is restricted to weaker coupling.
Extrapolating to a vanishing upper cut on $|X|$ brings $r_{\rm d}/2^{2-K}$ to within $1\%$ of unity for all $0.75\le K\le1.75$, both for the median over $\chi\ge32$ and for each run with $\chi\ge48$ that can be extrapolated (Appendix~\ref{app:window}).
The excess at large $K$ is expected from Eq.~\eqref{eq:flow}, since a growing $\gd$ pushes $K$ downward and the instantaneous rate $2^{2-K(\ell)}$ therefore increases across the fit window.
A synthetic flow with a running $K$ (Appendix~\ref{app:k175}) gives $r_{\rm d}/2^{2-K}=1.005$, $1.010$, and $1.018$ for $K=1.25$, $1.5$, and $1.75$, close to the measured $1.003$, $1.015$, and $1.027$.
It does not explain the deficit for $K\le1$.
At $K=1$ the deficit is accounted for by the fourth-order shift of the level at fixed $K$, which slows the growth of $|X|$; the exact free-fermion level reproduces the measured rate without free parameters (Appendix~\ref{app:k175}).
A term of the same form also removes the deficit at $K=0.6$ and $0.75$, but we have not computed its coefficient there.
The BKT point $K=1/2$ is treated in Sec.~\ref{sec:bkt}.

\begin{table}[tb]
  \caption{$O(2)$, $\chi=40$: fit window ($n_s=12$ to $n_e$), growth factor per step $r_{\rm d}$ with its range over fit windows (Sec.~\ref{sec:o2-rate}), its ratio to $2^{2-K}$, and the amplitude ratio of Sec.~\ref{sec:o2-amp} normalized by its prediction, with its range. $K=0.6$ is from an additional run with the same code and settings (Appendix~\ref{app:numerics}).}
  \label{tab:o2}
  \begin{ruledtabular}
    \begin{tabular}{lcccc}
      $K$  & window & $r_{\rm d}$                 & $r_{\rm d}/2^{2-K}$ & $\frac{A_J/A}{\alpha_J/(4K\alpha_{0,1})}$ \\
      \hline
      0.60 & 12--27 & 2.571 [2.571,2.623] & 0.974       & 0.995 [0.995,1.004]                       \\
      0.75 & 12--27 & 2.331 [2.329,2.362] & 0.980       & 1.005 [0.999,1.006]                       \\
      1.00 & 12--30 & 1.986 [1.985,1.996] & 0.993       & 1.003 [1.000,1.005]                       \\
      1.25 & 12--38 & 1.687 [1.686,1.689] & 1.003       & 0.998 [0.998,1.004]                       \\
      1.50 & 12--48 & 1.435 [1.425,1.435] & 1.015       & 1.003 [1.002,1.007]                       \\
      1.75 & 12--71 & 1.221 [1.210,1.221] & 1.027       & 1.059 [1.048,1.062]                       \\
    \end{tabular}
  \end{ruledtabular}
\end{table}

\subsection{Parameter-free amplitude ratio}
\label{sec:o2-amp}

For $0.75\le K\le1.5$ the relative amplitudes of the growth in two levels agree with the parameter-free prediction for $\Vd$ to within about $1\%$.
A growth rate fixes the dimension of a perturbation and the order at which it enters, but it does not identify the operator.
To identify the operator, we use the fact that a single coupling $\gd^2$ must produce the growth in every level, with relative weights set by the coefficients $\alpha_i(K)$.
We compare the charged level with the level of the currents $J,\Jb$ (dimension one, $e=m=0$, spin $\pm1$).
Its deviation $X_J(n)\coloneqq1-x_J(n)$ is insensitive to $\delta K$, since $\partial_Kx_J=0$, and receives no first-order umklapp contribution.
For $x_J$ we take the average of the two nearly degenerate levels among the lowest three in the neutral $C$-odd sector (the third is the $C$-odd combination of $e=\pm1$).
Since $\delta X=-4K\alpha_{0,1}\gd^2$ and $\delta X_J=-\alpha_J\gd^2$ to second order in $\gd$, the two growth amplitudes must satisfy
\begin{equation}
  \frac{A_J}{A}=\frac{\alpha_J(K)}{4K\alpha_{0,1}(K)},
  \label{eq:ampratio}
\end{equation}
with no adjustable parameter.
We obtain $A_J$ by fitting $X_J(n)$ in the same window with the same $r_{\rm d}$.

\begin{figure}[tb]
  \centering
  \includegraphics[width=\columnwidth]{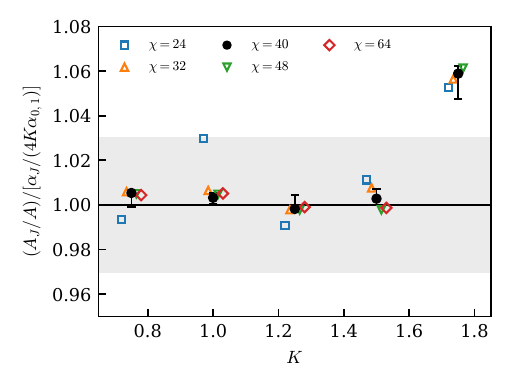}
  \caption{Ratio of the growth amplitudes in the current level and in the lowest charged level, normalized by the parameter-free second-order dimerization prediction~\eqref{eq:ampratio}, for $\chi=24$--$64$ ($O(2)$, standard tube). Error bars ($\chi=40$): range over fit windows. The agreement within about $1\%$ holds for the standard readout. Read with the wide tube, the ratio deviates from the prediction by a few percent unless the truncation of the additional readout step is relaxed (Sec.~\ref{sec:o2-amp}).}
  \label{fig:amp}
\end{figure}

For $0.75\le K\le1.5$ the measured ratio agrees with Eq.~\eqref{eq:ampratio} to within about $1\%$ for all $\chi\ge24$ (Fig.~\ref{fig:amp}; Table~\ref{tab:o2} for $\chi=40$).
The one exception, a deviation of $3\%$ at $\chi=24$ and $K=1$, is listed in Table~\ref{tab:chi}.
At the free-fermion point, where the prediction is $1/(4\ln2)=0.3607$, we measure $0.3618$.
The growth in both levels therefore comes from a single $C$-even, $m=0$ perturbation that acts at second order with the coefficients of $\Vd$, i.e., from the dimerization.
Because the ratio involves only $\gd^2$, it does not fix the sign of $\gd$.
We also checked the coefficients in Eq.~\eqref{eq:ampratio} by exact diagonalization and DMRG of the dimerized XXZ chain, independently of the Coulomb-gas integrals (Appendix~\ref{app:coefficients}).

At $K=1.75$ the ratio exceeds the prediction by $5$--$6\%$ for $\chi=24$--$48$.
Close to $K=2$ the running of $K$ induced by $\gd^2$ is important.
Including it raises the prediction to $1.071$ at $\chi=40$, close to the measured $1.059$, although an excess of about $3\%$ remains at the cut $|X|\le0.01$ for $\chi\ge48$ (Appendix~\ref{app:k175}).

The identification does not change when the same tensors are read with the wide tube.
The second-order deviation is then larger by the predicted factor $5^{2-K}$ to within about $6\%$, and the amplitude ratio agrees to within $0.5\%$ once the truncation of the extra coarse-graining step of this readout is relaxed to $2\chi$ (Appendix~\ref{app:numerics}).

\subsection{Level repulsion at \texorpdfstring{$K=1$}{K=1}}
\label{sec:k1}

At $K=1$ the spin-one states $J_{-1}|0,1\rangle$ and $|1,1\rangle$, of dimensions $1+1/(4K)$ and $K+1/(4K)$, are degenerate, and $\Vd$ connects them at first order.
Using $[J_n,V_{1,0}(z)]=\sqrt K\,z^nV_{1,0}(z)$, we find $|\langle1,1|\Vd(1)J_{-1}|0,1\rangle|=\sqrt{K/2}$.
With the factor $2\pi\gd$ of a first-order shift in the normalization of Eq.~\eqref{eq:coupling}, the matrix element between the two states is $\sqrt{2K}\,\pi\gd$.
The two-level approximation then predicts a splitting $\sqrt{(K-1)^2+8\pi^2K\gd^2}$, i.e., $2\sqrt2\pi|\gd|$ at $K=1$.
Because the splitting is linear in $\gd$, it measures $|\gd|$ independently of the second-order shifts used so far.

\begin{figure*}[tb]
  \centering
  \includegraphics[width=\textwidth]{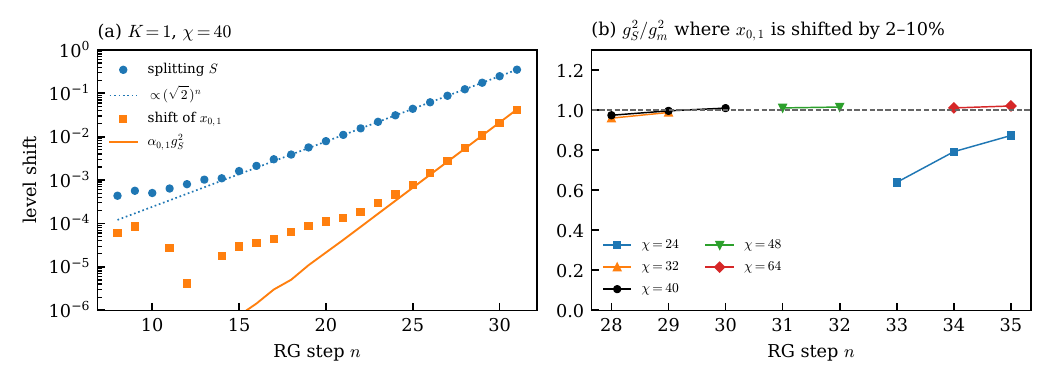}
  \caption{Level repulsion at $K=1$ ($O(2)$). (a) $\chi=40$: splitting $S$ in the $|m|=1$ sector (circles; defined in the text), growing as $(\sqrt2)^n$ (dotted), and the magnitude of the shift of $x_{0,1}$ from its floor (squares; at $n=10$ and $13$ it is below the range of the plot, at $n=13$ because the floor is the median) compared with $\alpha_{0,1}g_S^2$ predicted from $S$ (line). (b) Ratio $g_S^2/g_m^2$ of the dimerization couplings obtained from the first-order splitting and from the second-order shift of $x_{0,1}$, for $\chi=24$--$64$, where $x_{0,1}$ is shifted by $2$--$10\%$.}
  \label{fig:k1}
\end{figure*}

Let $S$ be the difference between the averages of the upper two and the lower two of the second to fifth levels of the $|m|=1$ sector, each of which contains one state of spin $+1$ and one of spin $-1$.
We define $g_S^2\coloneqq(S^2-S_{\rm floor}^2)/(8\pi^2K)$ and $g_m^2\coloneqq(x_{0,1}-x_{0,1}^{\rm floor})/\alpha_{0,1}$, with the floors taken as the medians over steps 8--14.
At $\chi=40$, $S$ grows by $1.412$ per step, close to the first-order rate $\sqrt2$ and clearly distinct from the second-order rate $2$ of $x_{0,1}$ (Fig.~\ref{fig:k1}).
The splitting leaves the floor already at $n\approx13$ ($\gd^2\sim10^{-8}$), while the deviation of $x_{0,1}$ is dominated by other contributions until $n\approx22$.
We compare the two in the window where $\gd^2$ dominates the deviation of $x_{0,1}$ ($2$--$10\%$).
There they agree, with $g_S^2/g_m^2=0.96$--$1.02$ for $\chi=32$--$64$.
The first-order splitting and the second-order shift thus give the same $\gd^2$, independently confirming the identification of Sec.~\ref{sec:o2-amp}.
At $\chi=24$ the ratio is lower, $0.64$--$0.87$, largely because the direct drift of $K$ (Sec.~\ref{sec:Kn}), which is largest at small $\chi$, adds to the shift of $x_{0,1}$ and inflates $g_m^2$ (Appendix~\ref{app:k175}).

\subsection{The BKT point \texorpdfstring{$K=1/2$}{K=1/2}}
\label{sec:bkt}

At the BKT point the levels follow the logarithmic flow of the $SU(2)_1$ theory for about 15 RG steps, after which the dimerization takes over as elsewhere on the critical line.
At $K=1/2$ ($\cv=2$) the critical line ends in a BKT transition~\cite{vanBeijeren1977}.
There the umklapp operator $\Vu$, of dimension $4K=2$, forms together with $J\Jb$ the marginally irrelevant current--current coupling of the $SU(2)_1$ theory, and the levels approach their CFT values only logarithmically~\cite{Cardy1986,Affleck1989}.
The four states of dimension $1/2$ split into an $SU(2)$ triplet $\{(0,\pm1)$, $C$-odd $e=\pm1\}$ and a singlet ($C$-even $e=\pm1$, i.e., $\Vd$).
To the order needed here~\cite{Affleck1989,AffleckCorr1990,Affleck1998,Barzykin1999}, their dimensions $x_t$ and $x_s$ are
\begin{equation}
  x_t=\tfrac12-\tfrac{\hat g}{4},\quad x_s=\tfrac12+\tfrac{3\hat g}{4},\quad \frac{d\hat g}{d\ell}=-\hat g^2-\tfrac12\hat g^3,
  \label{eq:bkt}
\end{equation}
where $\hat g>0$ is the marginally irrelevant coupling normalized by this equation.
In the combination $(x_s+3x_t)/4=1/2+O(\hat g^2)$ the term linear in $\hat g$ cancels.
Since the triplet members lie in different $O(2)$ sectors, their degeneracy also tests whether the truncation preserves the lattice $SU(2)$ symmetry of this point, which is not imposed.

\begin{figure*}[tb]
  \centering
  \includegraphics[width=\textwidth]{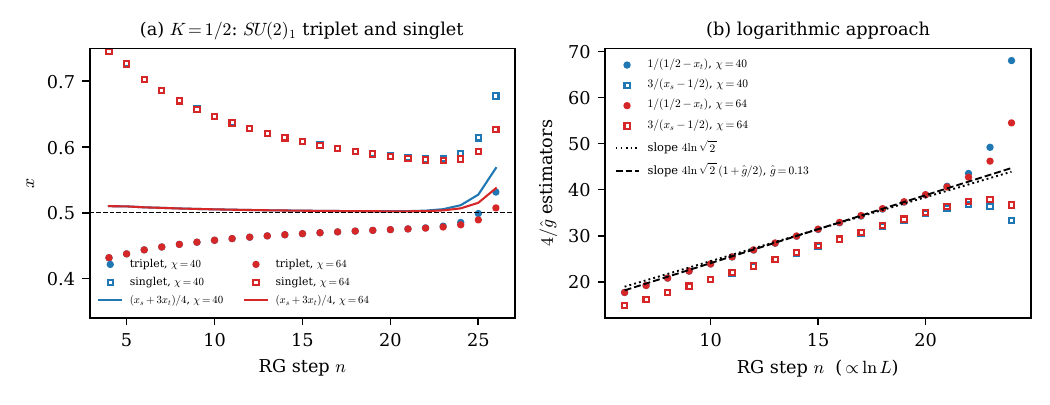}
  \caption{$O(2)$-symmetric LoopTNR at the BKT point $K=1/2$. (a) Triplet (average of the lowest $|m|=1$ and $C$-odd neutral levels; filled) and singlet ($C$-even $e=\pm1$; open) for $\chi=40$ and $64$, and the combination $(x_s+3x_t)/4$ (lines), in which the leading logarithm cancels. (b) $1/(1/2-x_t)$ and $3/(x_s-1/2)$, which equal $4/\hat g$ to leading order, compared with the slopes $4\ln\sqrt2$ (dotted) and $4\ln\sqrt2\,(1+\hat g/2)$ (dashed) predicted by Eq.~\eqref{eq:bkt}; both lines pass through the $\chi=64$ triplet point at $n=15$, and the dashed one uses $\hat g=0.13$, the value at that point.}
  \label{fig:bkt}
\end{figure*}

The charged doublet and the $C$-odd neutral level, which together form the triplet, stay degenerate to $2\times10^{-5}$ at $\chi=64$ for $n\le22$ (Fig.~\ref{fig:bkt}).
The triplet approaches $1/2$ from below and the singlet from above.
At $n=20$ ($\chi=64$) the two give $\hat g=4(1/2-x_t)=0.103$ and $\tfrac43(x_s-1/2)=0.114$, which are equal within $O(\hat g^2)$.
The combination $(x_s+3x_t)/4-1/2=2.1\times10^{-3}$ is an order of magnitude smaller than $\hat g$.

How fast the levels approach their limit is sensitive to the two-loop term.
With $\ln L=n\ln\sqrt2+{\rm const}$, Eq.~\eqref{eq:bkt} predicts that $1/(1/2-x_t)=4/\hat g$ grows by $4\ln\sqrt2\,(1+\hat g/2)$ per step.
Over $n=8$--$22$ we measure $1.49$--$1.55$ per step for $\chi=40$, $48$, and $64$ [Fig.~\ref{fig:bkt}(b)], closer to the two-loop value than to the leading-logarithm value $1.386$.
Over this range the two-loop value decreases from $1.52$ to $1.45$, as $\hat g$ decreases from about $0.19$ to $0.09$.
The upper values, at $\chi=40$ and $64$, come from $n=21$ and $22$, where the onset of the dimerization already raises $1/(1/2-x_t)$.
Over $n=8$--$20$ all three give $1.49$--$1.50$, against $1.48$ for the two-loop value averaged over these steps.
The remaining difference is of the order of the neglected $O(\hat g^2)$ terms.
Up to $n=22$ the triplet levels for $\chi=40$ and $64$ agree to $5\times10^{-4}$, and the singlet levels to $1.5\times10^{-3}$.
The logarithmic approach, which we follow here over $15$ RG steps, therefore changes little between $\chi=40$ and $64$, and we attribute it to the lattice model rather than to the truncation.

Beyond $n\approx22$ the dimerization takes over, and the window ends at $n_e=26$--$29$.
Because $\Vd$ is the $SU(2)$ singlet, its dimension runs as $x_s=1/2+3\hat g/4$, and its second-order growth factor is $2^{3/2-3\hat g/4}$, on average $0.944$--$0.947$ times $2^{3/2}$ over the fit window.
In addition, the background follows $X=\hat g/2$ instead of a straight line.
With both effects included, the measured growth factor is $0.958$--$0.985$ times the expectation for $\chi=32$--$64$, a value that depends on the treatment of the background and of the running rate (Appendix~\ref{app:k175}).

\subsection{Motion along the critical line}
\label{sec:Kn}

Before the dimerization grows, the spectrum is that of a compact boson whose Luttinger parameter drifts slowly, by less than $2\times10^{-4}$ per step for $\chi\ge32$.
We distinguish three shifts of $K$: the \emph{direct} drift caused by the truncation, a shift that we \emph{inject} by hand to test the estimator, and the drift \emph{induced} by the growing dimerization through Eq.~\eqref{eq:flow}.
We find no amplification or damping of the injected shift within the errors and that the direct drift is consistent with an accumulated marginal error.
At $\chi=40$ the induced drift at the end of the fit window is several times larger than the direct one.

To follow the position on the critical line, we determine at each step an effective Luttinger parameter $K_n$ and the common scale error $\eta$ from two levels, assuming $x_i=(1+\eta)\,x_i^{\rm CFT}(K_n)$.
We use the lowest charged level, $x_{0,1}^{\rm CFT}=1/(4K)$, and the average of the $C$-even and $C$-odd combinations of $e=\pm1$, $x_e^{\rm CFT}=K$, in which their first-order umklapp splitting cancels.
This gives
\begin{equation}
  K_n=\frac12\sqrt{\frac{x_e}{x_{0,1}}},\qquad 1+\eta=2\sqrt{x_e\,x_{0,1}} .
  \label{eq:Kn}
\end{equation}
We define the floor window as the steps $n\ge12$ after the dimension-four terms have decayed and before the fitted dimerization term exceeds $10^{-4}$; the run at $K=1.75$ and $\chi=64$, whose charged level does not reach $|X|=0.1$ within 80 steps, is excluded.
In this window the levels not used in the estimate, $x_{0,2}$, $x_{0,3}$, and $x_J$, agree with their predictions from $K_n$ and $\eta$ to within $2\times10^{-3}$ for $\chi\ge32$, with one exception (Appendix~\ref{app:k175}), while $|\eta|\lesssim10^{-3}$.
At $K=1$, where $x_J$ and the $C$-odd combination of $e=\pm1$ are degenerate at the CFT point, the level assigned to $x_J$ is the one already used in $x_e$, so there $x_J$ is not an independent check.
The spectrum is thus that of a compact boson at a slowly moving $K_n$.
The second marginal direction, the lattice anisotropy coupled to $T,\bar T$, is seeded only weakly (Appendix~\ref{app:k175}).

\begin{figure*}[tb]
  \centering
  \includegraphics[width=\textwidth]{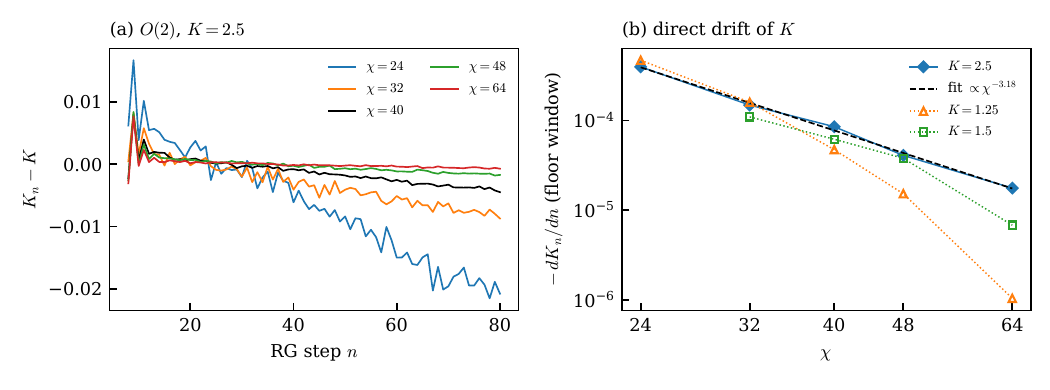}
  \caption{Motion along the critical line ($O(2)$). (a) $K_n-K$ from Eq.~\eqref{eq:Kn} at $K=2.5$, where no relevant perturbation grows, for $\chi=24$--$64$. (b) Direct drift $-dK_n/dn$ in the floor window (steps $n\ge12$ before the fitted dimerization term exceeds $10^{-4}$) versus $\chi$; filled: $K=2.5$ (whole run, with a power-law fit); open: $K=1.25$ and $1.5$ (shorter floor window).}
  \label{fig:drift}
\end{figure*}

The direct drift is the slope $dK_n/dn$ in the floor window, which is proportional to the source $\sigma_K$ of Eq.~\eqref{eq:sourceflow} before the dimerization is large enough to induce a drift.
At $K=2.5$, where no dimerization grows, the whole run can be used.
There $K_n$ decreases linearly, by $8.5\times10^{-5}$ per step at $\chi=40$, and $|dK_n/dn|\propto\chi^{-3.2}$ (Fig.~\ref{fig:drift}).
For $K<2$ the floor window is shorter, but the drift is of the same order and also decreases with $\chi$, roughly as $\chi^{-6}$ at $K=1.25$.

To test both the estimator and the marginality of the drift, we inject a known shift of $K$ by comparing runs at $K\pm10^{-3}$ with the reference runs at $K=1.25$ and $2.5$ ($O(2)$, $\chi=40$).
To first order this shift is the exactly marginal $J\Jb$ deformation.
The estimator~\eqref{eq:Kn} reproduces the injected shift, with a response of $1.002\pm0.006$ at $K=1.25$ ($n=8$--$24$) and $0.987\pm0.019$ at $K=2.5$ ($n=8$--$70$).
Its slope per step, $(+1.8\pm1.3)\times10^{-3}$ and $(-1.8\pm1.0)\times10^{-3}$, is within two standard errors of zero, so no amplification or damping is resolved (Appendix~\ref{app:k175}).
Written as a growth factor $2^{y_K/2}$ per step, it has $|y_K|\lesssim0.01$, compared with $2-K=0.75$ for the dimerization at $K=1.25$.
The direct drift of $K_n$ in the floor window is therefore consistent with an accumulated marginal source rather than an earlier error amplified along the critical line.
This assumes that a common drift of the readout is negligible.
Alternative estimators of $K$ support this assumption at $K=2.5$ but not uniformly for $K<2$ (Appendix~\ref{app:k175}).

The drift induced by the dimerization is larger.
Table~\ref{tab:dK} compares the direct drift accumulated up to the end of the fit window, $|dK_n/dn|\,n_e$, with the induced shift $|\delta K|\simeq\pi^2K^2\gd^2/(2-K)$, estimated from Eq.~\eqref{eq:flow} at fixed $K$ with $\gd^2$ at $n_e$ from the growth fit.
At $\chi=40$ the induced shift is larger by a factor of $11$--$12$ for $K\ge1.25$ and of $5$--$6$ at $K=1$, where it depends on the choice of levels (Appendix~\ref{app:k175}).
At $K=0.75$ the direct drift is close to zero and of opposite sign ($+4.7\times10^{-6}$ per step), so the large ratio there carries little information.
Since the induced shift at $n_e$ is set by the breakdown criterion, the ratio tends to grow with $\chi$ as the direct drift decreases.
At $\chi=24$ and $32$ it is only $0.6$--$12$ (Appendix~\ref{app:k175}).

\begin{table}[tb]
  \caption{Shift of $K$ accumulated up to the end of the fit window at $\chi=40$. Direct: slope $|dK_n/dn|$ in the floor window times $n_e$. Induced: $\pi^2K^2\gd^2(n_e)/(2-K)$ from Eq.~\eqref{eq:flow} at fixed $K$, with $\gd^2(n_e)$ from the growth fit. At $K=0.75$ the floor-window slope is positive and close to zero.}
  \label{tab:dK}
  \begin{ruledtabular}
    \begin{tabular}{rccc}
      $K$  & direct $|\delta K|$ & induced $|\delta K|$ & ratio \\
      \hline
      0.75 & $1.3\times10^{-4}$  & $0.003$              & 22    \\
      1.00 & $1.2\times10^{-3}$  & $0.008$              & 6     \\
      1.25 & $1.8\times10^{-3}$  & $0.019$              & 11    \\
      1.50 & $3.0\times10^{-3}$  & $0.037$              & 12    \\
      1.75 & $6.3\times10^{-3}$  & $0.071$              & 11    \\
    \end{tabular}
  \end{ruledtabular}
\end{table}

The induced drift can also be extracted from the spectrum within the fit window.
For $K=1.25$ and $1.5$, at the end of the window, it reaches $75$--$93\%$ of the shift $-\pi^2K^2\gd^2/(2-K)$ that Eq.~\eqref{eq:flow} predicts at fixed $K$ (Appendix~\ref{app:k175}).
This extraction uses the perturbative coefficients and the coupling $\gd^2(n)$ from the growth fit of the charged level.
It is therefore a consistency check of Eq.~\eqref{eq:flow} within the perturbative description, not an independent measurement.

\section{Reduced symmetry: \texorpdfstring{$U(1)$}{U(1)} only and \texorpdfstring{$C$}{C} only}
\label{sec:reduced}

We now impose only part of $O(2)$: only $U(1)$ (Secs.~\ref{sec:u1} and \ref{sec:u1-phase}) or only charge conjugation $C$ (Sec.~\ref{sec:conly}).

\subsection{\texorpdfstring{$U(1)$}{U(1)} only: current perturbation}
\label{sec:u1}

With only $U(1)$ imposed, the levels of opposite charge split, and the splitting grows by $\sqrt2$ per step while its sign pattern rotates by $135^\circ$ per step, as expected for a first-order response to the $U(1)$ current.
The dimerization is present as well and grows as in the $O(2)$ runs.

Since $C$ is no longer protected, we decompose the lowest levels $x_{0,\pm q}(n)$ into
\begin{equation}
  \underline{x}_q\coloneqq\tfrac12(x_{0,+q}+x_{0,-q}),\qquad \Oq_q\coloneqq\tfrac12(x_{0,+q}-x_{0,-q}).
  \label{eq:EO}
\end{equation}
By Eq.~\eqref{eq:current-shift}, the current contributes $\Oq_q=(\pi q/\sqrt K)g_-$ at first order, with a magnitude growing by $\sqrt2$ and a phase rotating by $135^\circ$ per step (Table~\ref{tab:rates}).
The fields $\Vd$ and $\Vs$, in contrast, shift $\underline{x}_q$ by $\alpha_{0,q}(\gd^2+g_{\mathrm s}^2)$ and do not contribute to $\Oq_q$ at second order (Appendix~\ref{app:u1checks}).

We read these levels in the \emph{two-level window}, the contiguous range of steps from $n=15$ in which both charged levels are within $10\%$ of $1/(4K)$ and $|\mathsf{c}-1|<0.2$ (Appendix~\ref{app:numerics}).
Within these windows the lowest charged levels are real and split in modulus, and, for $1\le K\le1.75$, the $(\pm1,0)$ pair has equal moduli and opposite phases, as expected from Sec.~\ref{sec:cpt}.

\begin{figure*}[tb]
  \centering
  \includegraphics[width=\textwidth]{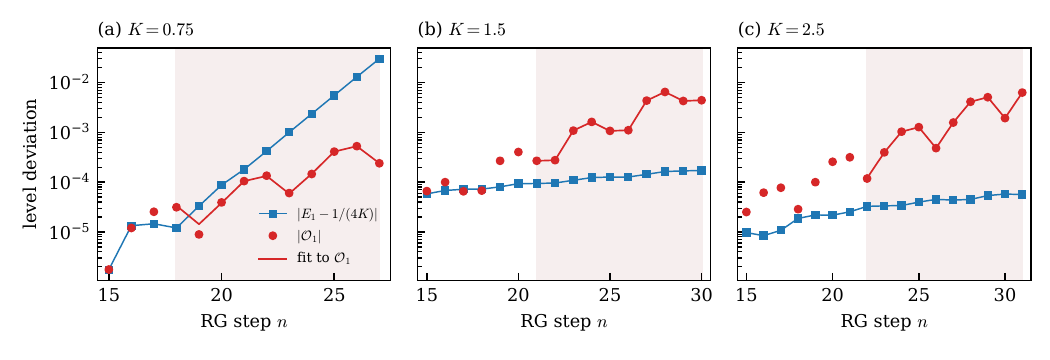}
  \caption{$U(1)$-symmetric LoopTNR at $\chi=40$: deviation of the charge-averaged level, $|\underline{x}_1-1/(4K)|$ (squares; background not subtracted), and the charge-odd splitting $|\Oq_1|$ (circles), for (a) $K=0.75$, (b) $K=1.5$, and (c) $K=2.5$. The shaded region is the fit window, and the line shows the absolute value of the fit~\eqref{eq:u1fit} to the signed $\Oq_1$, evaluated at integer steps.}
  \label{fig:u1flow}
\end{figure*}

Figure~\ref{fig:u1flow} shows $|\underline{x}_1-1/(4K)|$ and $|\Oq_1|$ for $K=0.75$, $1.5$, and $2.5$ at $\chi=40$.
The odd part $\Oq_1$ grows with an oscillating sign at all three values, including $K=2.5$, where $\Vd$ and $\Vs$ are irrelevant.
The even part grows rapidly only at $K=0.75$, with the same growth factor as in the $O(2)$ case ($r_{\rm d}/2^{2-K}=0.976$--$0.985$ for $\chi=32$--$64$).
For $K\le1$ the even part ends the window, at the same step as in the $O(2)$ runs.
For $K\ge1.25$ the window is instead closed by the current, whose first-order growth by $\sqrt2$ per step competes with the second-order dimerization growth $2^{2-K}\le2^{3/4}$ (Appendix~\ref{app:u1checks}).

We fit the signed $\Oq_1(n)$ with
\begin{equation}
  \Oq_1(n)=r_J^{\,\nu}\left[u\cos(\omega\nu)+v\sin(\omega\nu)\right],\qquad \nu\coloneqq n-n_e,
  \label{eq:u1fit}
\end{equation}
with the growth factor $r_J$ and the rotation angle $\omega$ as free parameters ($0<\omega<\pi$; a real signal does not distinguish $\omega$ from $-\omega$) and the amplitudes $u,v$ from linear least squares.
The fit uses the last ten steps of the two-level window.

\begin{figure}[tb]
  \centering
  \includegraphics[width=\columnwidth]{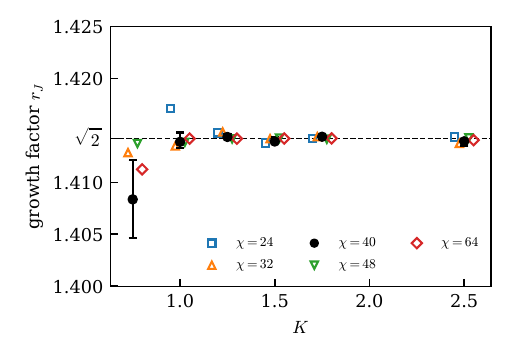}
  \caption{Growth factor per step $r_J$ of the charge-odd splitting $\Oq_1$ for $U(1)$ symmetry and $\chi=24$--$64$ (shifted horizontally), with $r_J$ and $\omega$ both free. Error bars ($\chi=40$): range over fit windows of 8, 10, and 12 points with the end point shifted by 0, 1, and 2 steps (not a statistical confidence interval). Dashed line: first-order current prediction $\sqrt2$.}
  \label{fig:u1gamma}
\end{figure}

Figure~\ref{fig:u1gamma} and Table~\ref{tab:u1} show the results for six values of $K$.
For $K\ge1$, $r_J$ agrees with $\sqrt2$ to within $0.03\%$ and $\omega$ with $135^\circ$ to within $0.02^\circ$ at $\chi=40$.
This is what one expects for a first-order response to a spin-one operator of dimension one, i.e., the $U(1)$ current.
At $K=1.5$ the second-order rate of $\Vd$ and $\Vs$ is also $2^{2-K}=\sqrt2$, but these operators explain neither the oscillation nor the persistence of the growth for $K>2$.

A uniform current shifts only the zero modes and gives $\Oq_q\propto q$ to all orders, which holds to within $4.2\%$ for $K\ge1$.
At $K=0.75$, where the even part is large, $r_J$ is $0.4\%$ below $\sqrt2$ and the proportionality is violated by $24\%$, indicating contributions beyond the leading current response (Appendix~\ref{app:u1checks}).

The real part of $g_J$ couples to $\int dy\,(J-\Jb)\propto\int dy\,\partial_y\phi$, i.e., to the $U(1)$ charge, so it acts as a chemical potential, the analog of an external field in the six-vertex model.
It shifts the charged levels linearly in the charge, leaves the charge-averaged levels $\underline{x}_q$ unaffected, and neither opens a gap nor moves $K$.
For $K\ge1.25$ the two-level window closes because the charged levels fail the readout criterion.
This alone does not imply that the run leaves the critical line.
At $\chi=40$ the charge-averaged levels $\underline{x}_1$ and $\underline{x}_2$ and the neutral levels stay within $0.5\%$ of their CFT values for 8 and 12 steps after the window closes (to $n=38$ at $K=1.5$ and $n=43$ at $K=2.5$), while the charge splitting exceeds the levels themselves.
Over the same steps the central charge read from $\lambda_0$ becomes unreliable, presumably because the current also affects the normalization of the tensors.
The window criterion $|\mathsf c-1|<0.2$ should be read with this in mind.

A readout with half the circumference halves $\Oq_1$, as expected for a first-order response to an operator of dimension one (Appendix~\ref{app:u1checks}).

\begin{table}[tb]
  \caption{$U(1)$, $\chi=40$: fit window, growth factor per step $r_J$ of the charge-odd splitting with its range over fit windows (as in Fig.~\ref{fig:u1gamma}), and rotation angle $\omega$ per step.}
  \label{tab:u1}
  \begin{ruledtabular}
    \begin{tabular}{lccc}
      $K$  & window & $r_J$                & $\omega$ (deg) \\
      \hline
      0.75 & 18--27 & 1.4084 [1.4047, 1.4121] & 135.16         \\
      1.00 & 21--30 & 1.4139 [1.4134, 1.4148] & 135.01         \\
      1.25 & 20--29 & 1.4144 [1.4143, 1.4147] & 135.01         \\
      1.50 & 21--30 & 1.4139 [1.4139, 1.4144] & 134.99         \\
      1.75 & 26--35 & 1.4144 [1.4144, 1.4144] & 135.00         \\
      2.50 & 22--31 & 1.4140 [1.4135, 1.4140] & 135.00         \\
    \end{tabular}
  \end{ruledtabular}
\end{table}

\subsection{Imaginary part of the current coupling}
\label{sec:u1-phase}

The complex coupling $g_J$ obtained from the charge splitting also predicts the phases of the eigenvalues of the neutral states $(\pm1,0)$, with no free parameter.
For $K\ge1$ the measured phases agree with this prediction to within $3\%$.
With $g_-=2\operatorname{Re}g_J$ and $g_J\propto(\sqrt2e^{3\pi i/4})^n$, the fit~\eqref{eq:u1fit} determines the complex current coupling up to complex conjugation,
\begin{equation}
  g_J(n)=\frac{\sqrt K}{2\pi}\,(u-iv)\,r_J^{\nu}e^{i\omega\nu},\qquad \nu\coloneqq n-n_e .
  \label{eq:gJ}
\end{equation}
Its magnitude at the end of the window, $|g_J(n_e)|=\sqrt K\,(u^2+v^2)^{1/2}/(2\pi)$, lies between $10^{-4}$ and $6\times10^{-3}$ for the runs studied.
Equation~\eqref{eq:gJ} then predicts a second observable without free parameters.
Since $\operatorname{Im}g_J$ shifts the neutral states $(\pm1,0)$ by $\pm4\pi i\sqrt K\operatorname{Im}g_J$ (Sec.~\ref{sec:cpt}), their transfer-matrix eigenvalues acquire phases
\begin{equation}
  \bigl|\arg(\lambda_{\pm1,0}/\lambda_0)\bigr|=2\pi K\,r_J^{\nu}\,\bigl|u\sin\omega\nu-v\cos\omega\nu\bigr|
  \label{eq:phase-pred}
\end{equation}
in the readout with $\operatorname{Im}\tau=1/2$, while their moduli are unaffected at first order.

\begin{figure}[tb]
  \centering
  \includegraphics[width=\columnwidth]{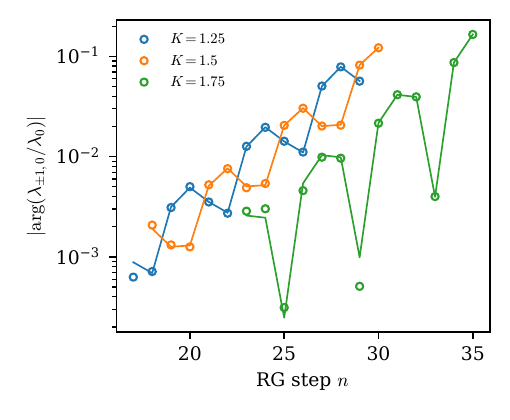}
  \caption{$U(1)$, $\chi=40$, $K=1.25$, $1.5$, and $1.75$: eigenvalue phases of the neutral levels with dimension $x\simeq K$ (circles) and the parameter-free prediction~\eqref{eq:phase-pred} from the fit of the charge splitting $\Oq_1$ (lines, evaluated at integer steps), over the fit window and the preceding steps.}
  \label{fig:u1phase}
\end{figure}

Figure~\ref{fig:u1phase} compares the measured phases of the neutral levels with $x\simeq K$ to this prediction, which is obtained from the fit of $\Oq_1$ at $\chi=40$.
The prediction reproduces the growth and the oscillation, including the near zeros.
Over the fit windows the ratio of measured to predicted phases is $0.97$--$1.002$, with a correlation coefficient $\ge0.999$, for $K=1$, $1.25$, $1.5$, and $1.75$ and all $\chi=24$--$64$.
At $K=1$ the two current levels, which are degenerate with $(\pm1,0)$ and carry no phase, are excluded by selecting only the complex-conjugate pair.
A single complex coupling $g_J$ rotating by $135^\circ$ per step thus accounts for both the charge splitting and the phases.
This also confirms the reality condition $g_{\Jb}=-g_J^*$ of Sec.~\ref{sec:cpt}.
At $K=2.5$, where dimerization is irrelevant, the $(\pm1,0)$ levels lie above the eight levels of the standard readout.
We therefore read the saved tensors with 24 neutral levels and obtain a ratio of $0.998$ at $\chi=40$, with a correlation coefficient $\ge0.9999$ (Appendix~\ref{app:u1checks}).
At $K=0.75$ a mixing of $(1,0)$ and $(-1,0)$, from the umklapp term early in the window and from the second-order dimerization later, usually exceeds the predicted imaginary shift and keeps the two levels near $x=K$ real.
At $\chi=32$, where $g_J$ is larger, they become complex, with phases close to the prediction, and at $\chi=48$ they are complex at part of the steps (Appendix~\ref{app:u1checks}).

\subsection{\texorpdfstring{$C$}{C} only: charge-two vertex operator}
\label{sec:conly}

With only $C$ imposed, the leading $U(1)$-breaking error is the charge-two vertex operator $V_{0,\pm2}$, although the more relevant $V_{0,\pm1}$ is allowed as well.
We impose $C$ by transforming each bond from the arrow basis to the eigenbasis of arrow reversal and using $\mathbb Z_2$-graded tensors.
Code and parameters are otherwise those of Secs.~\ref{sec:o2} and \ref{sec:u1} ($K=0.75$--$2.5$, $\chi=24$, $32$, $40$, up to 40 steps, 16 levels per $C$ sector).
The primaries $(0,\pm1)$ then appear as one $C$-even and one $C$-odd level, both at $1/(4K)$, and their half splitting $\Delta x\coloneqq\tfrac12(x_{\rm even}-x_{\rm odd})$ measures the mixing of $(0,1)$ and $(0,-1)$.
Since $V_{0,\pm2}$ connects these two degenerate states at first order, $\Delta x\propto g_{0,2}$ grows by $2^{1-1/(2K)}$ per step, without oscillation because the operator has spin zero.
The operator $V_{0,\pm1}$ connects them only at second order, through the neutral sector, which would give a factor $2^{2-1/(4K)}$, i.e., $3.2$--$3.7$ for the values of $K$ studied.
The dimerization $\Vd$ preserves $U(1)$ and does not split them.

\begin{figure*}[tb]
  \centering
  \includegraphics[width=\textwidth]{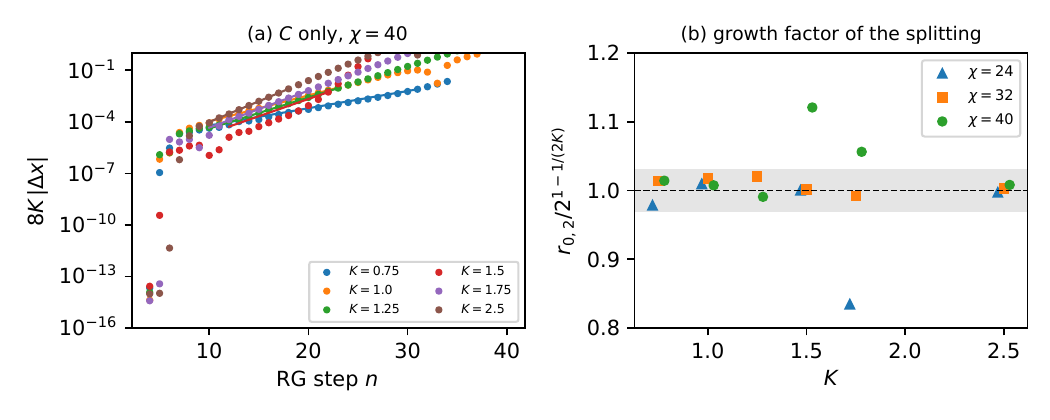}
  \caption{LoopTNR with only $C$ imposed. (a) Splitting of the $C$-even and $C$-odd levels of the charge-$\pm1$ primaries at $\chi=40$ (dots), with the first-order $V_{0,\pm2}$ rate $2^{1-1/(2K)}$ drawn over each fit window (lines). (b) Fitted growth factor $r_{0,2}$ of the splitting divided by $2^{1-1/(2K)}$ for $\chi=24$, $32$, $40$ (shifted horizontally); the shaded band indicates $\pm3\%$. The second-order $V_{0,\pm1}$ rate would correspond to $2^{1+1/(4K)}=2.1$--$2.5$ on this scale.}
  \label{fig:conly}
\end{figure*}

The splitting $\Delta x$ stays at the round-off level ($\lesssim10^{-14}$) as long as the bond dimension is below $\chi$.
Within one or two steps after the truncation to $\chi$ sets in ($n=5$--$6$), it jumps by eight to ten orders of magnitude, and it then grows exponentially with a constant sign [Fig.~\ref{fig:conly}(a)].
We fit $\ln|\Delta x|$ linearly over the contiguous steps $n\ge10$ in which $8K|\Delta x|$ lies between $10^{-5}$ and $10^{-2}$ (Appendix~\ref{app:numerics}).
For $\chi=32$ and $40$, the fitted growth factor $r_{0,2}$ of $\Delta x$ divided by $2^{1-1/(2K)}$ is $0.991$--$1.021$ at all six values of $K$, except in two cases at $\chi=40$ [Fig.~\ref{fig:conly}(b)].
These are $K=1.75$, where the factor is $1.06$, and $K=1.5$, where it is $1.12$ because the splitting stays near $8K|\Delta x|\sim10^{-6}$ until $n\approx14$ and then accelerates as the run approaches breakdown.
No component with the second-order rate of $V_{0,\pm1}$ is seen in the growth of the splitting.
The $U(1)$-breaking error seeded at the first truncations is thus $V_{0,\pm2}$, even though $V_{0,\pm1}$ is also permitted by $C$ and is more relevant.
The origin of this selection remains unclear.
Cutting through the degenerate $C$ doublets $\{|q\rangle,|{-q}\rangle\}$ of a bond does not account for it (Appendix~\ref{app:conly}).
The operator $V_{0,\pm1}$ is still generated, but at a smaller level.
In 6 of the 18 runs, a simultaneous fit of the mean and the splitting of the two levels requires its second-order term near the end of the weak-coupling window, with the ratio of contributions predicted for $V_{0,\pm1}$.
A deliberate injection with a transverse field confirms this signature (Appendix~\ref{app:conly}).

With the window criterion of Sec.~\ref{sec:u1}, the $C$-only runs at $\chi=40$ and $K\le1$ leave the weak-coupling regime at about the same step as the $O(2)$ runs.
For $K\ge1.25$ they fail earlier: at $K=1.25$ with $n_e=30$ compared with $38$ for $O(2)$, and by a factor of two or more at $K\ge1.5$, with $n_e=24$ and $26$ at $K=1.5$ and $1.75$ compared with $48$ and $71$ (Appendix~\ref{app:conly}).
The first-order splitting due to $V_{0,\pm2}$ grows as $L^{2-1/K}$, and the second-order dimerization contribution as $L^{2(2-K)}$.
The former grows faster only for $K>(1+\sqrt3)/2\approx1.37$.
At $K=1.25$, the earlier breakdown of the $C$-only run must therefore come from a larger seed of $V_{0,\pm2}$, not from faster growth.

\section{Bond-dimension dependence}
\label{sec:chi}

Changing the bond dimension changes the seed, while the operator that grows and its growth rate stay the same.
We repeated the analysis of Sec.~\ref{sec:o2} and of the $U(1)$-only runs of Sec.~\ref{sec:u1} for $\chi=16$--$64$, and the $O(2)$ analysis also for $\chi=96$ at $K=1$ and $1.5$, with identical code and parameters (Table~\ref{tab:chi}).
The $C$-only runs of Sec.~\ref{sec:conly} cover $\chi=24$--$40$.

Where reliable fits are available, the growth rate and the amplitude ratio depend only weakly on $\chi$.
At fixed $K$, $r_{\rm d}/2^{2-K}$ varies with $\chi$ by about $1\%$ or less ($0.976$--$0.987$ at $K=0.75$, $0.992$--$0.999$ at $K=1$, $1.002$--$1.003$ at $K=1.25$ for $\chi\ge32$, and $1.012$--$1.015$ at $K=1.5$), and the normalized amplitude ratio is $0.991$--$1.011$ for $0.75\le K\le1.5$, except $1.030$ at $\chi=24$ and $K=1$.
At $K=1.75$ the rate ($1.027$--$1.029$) and the amplitude ratio ($1.05$--$1.06$) are equally stable for $\chi=24$--$48$; at $\chi=64$ the window exceeds the 80 steps of the run.
The $U(1)$ runs behave in the same way, with $r_J$ within $0.4\%$ of $\sqrt2$ and the rotation per step within $0.2^\circ$ of $135^\circ$ for all fitted runs with $\chi\ge24$.
At $K=0.75$ and $\chi=24$ the window is too short for the fit.
The $\chi=16$ runs are too short for a reliable fit and are not included in these ranges.

\begin{figure}[tb]
  \centering
  \includegraphics[width=\columnwidth]{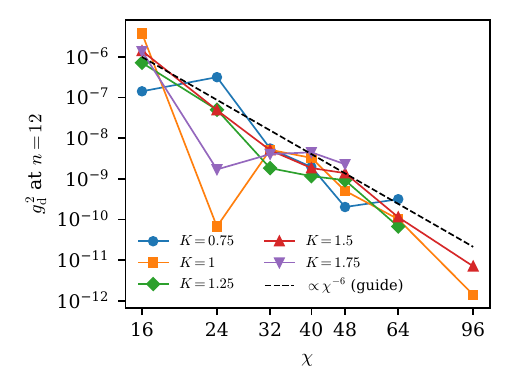}
  \caption{$O(2)$: seed $\gd^2(12)$, the dimerization coupling at $n=12$ obtained by extrapolating the growth term of the charged level with the fitted growth factor, versus $\chi$ for $K=0.75$--$1.75$ ($\chi=16$--$64$; $\chi\le48$ at $K=1.75$) and for $K=1$ and $1.5$ at $\chi=96$. Dashed line: $\chi^{-6}$, close to the separate fits at $K=1.25$ and $1.5$, where the fluctuations are smallest (exponents $-6.0\pm1.1$ and $-6.2\pm0.4$), as a guide; a common fit for $\chi\ge24$ gives $\chi^{-5.1\pm0.8}$, and $\chi^{-4.3\pm1.0}$ without $\chi=96$ (Sec.~\ref{sec:chi}).}
  \label{fig:seed}
\end{figure}

The bond dimension instead controls the size of the seed $\gd^2(12)$ (Fig.~\ref{fig:seed}).
From $\chi=16$ to $64$ it decreases by about four orders of magnitude for $K=1$--$1.5$ and by about three at $K=0.75$, and the breakdown moves to correspondingly later steps.
The decrease is far from monotonic, however, and at $K=1.75$ the seed shows no systematic decrease for $\chi=24$--$48$.
Individual runs deviate from the trend by up to two orders of magnitude.
For instance, the run at $K=1$ and $\chi=24$ has a smaller seed than those at $\chi=32$--$64$.
A common power-law fit to all $K=0.75$--$1.75$ for $\chi\ge24$, with separate amplitudes for each $K$, gives $\gd^2\propto\chi^{-5.1\pm0.8}$ with a residual scatter of $0.7$ decades (Fig.~\ref{fig:seed}).
Round-off differences between runs branched with identical settings from a common checkpoint change the seed by at most about $3\%$ (\nolinkurl{runs/additional/} in the data repository), so the fluctuations originate in the algorithm itself, i.e., in its truncation and loop optimization.
The companion paper will show that they depend strongly on the loop optimization~\cite{Paper1b}.

\begin{table}[tb]
  \caption{$O(2)$: end of the weak-coupling window $n_e$, growth factor per step relative to $2^{2-K}$, amplitude ratio normalized by Eq.~\eqref{eq:ampratio}, and the seed $\gd^2(12)$, for $\chi=16$--$96$.}
  \label{tab:chi}
  \begin{ruledtabular}
    \begin{tabular}{rrcccc}
      $K$  & $\chi$ & $n_e$ & $r_{\rm d}/2^{2-K}$ & ratio & $\gd^2(12)$         \\
      \hline
      1.00 & 16     & 20    & 0.967       & 0.904 & $3.8\times10^{-6}$  \\
      1.00 & 24     & 35    & 0.999       & 1.030 & $6.8\times10^{-11}$ \\
      1.00 & 32     & 29    & 0.996       & 1.006 & $5.2\times10^{-9}$  \\
      1.00 & 40     & 30    & 0.993       & 1.003 & $3.3\times10^{-9}$  \\
      1.00 & 48     & 32    & 0.995       & 1.005 & $5.1\times10^{-10}$ \\
      1.00 & 64     & 35    & 0.992       & 1.005 & $1.0\times10^{-10}$ \\
      1.00 & 96     & 41    & 0.993       & 1.003 & $1.4\times10^{-12}$ \\
      1.50 & 16     & 29    & 1.032       & 0.983 & $1.4\times10^{-6}$  \\
      1.50 & 24     & 39    & 1.013       & 1.011 & $4.9\times10^{-8}$  \\
      1.50 & 32     & 45    & 1.014       & 1.008 & $5.1\times10^{-9}$  \\
      1.50 & 40     & 48    & 1.015       & 1.003 & $1.9\times10^{-9}$  \\
      1.50 & 48     & 49    & 1.014       & 0.998 & $1.4\times10^{-9}$  \\
      1.50 & 64     & 56    & 1.012       & 0.999 & $1.1\times10^{-10}$ \\
      1.50 & 96     & 64    & 1.012       & 0.998 & $7.1\times10^{-12}$ \\
    \end{tabular}
  \end{ruledtabular}
\end{table}

\section{Summary and discussion}
\label{sec:discussion}

We have analyzed finite-bond-dimension LoopTNR on the critical line of the six-vertex model with symmetry-resolved transfer-matrix spectroscopy and conformal perturbation theory.
The truncation error acts as a perturbation by an operator that the preserved symmetries allow.
With $O(2)$ imposed this operator is the dimerization, with only $U(1)$ imposed the $U(1)$ current appears in addition, and tensors with only $C$ generate $V_{0,\pm2}$ as the leading $U(1)$-breaking error (Table~\ref{tab:summary}).
Before the dimerization grows, the spectrum is that of a compact boson at a slowly moving $K_n$, and at the BKT point the marginal flow is reproduced over 15 RG steps.
While the operators and their growth rates do not depend on $\chi$, the seed varies irregularly with $\chi$ (Fig.~\ref{fig:seed}).
For LoopTNR, the finite-$\chi$ errors at criticality can thus be identified operator by operator from the symmetries that the coarse-graining map actually preserves.

Which of the permitted operators dominates is not decided by the scaling dimension alone.
The size of the coupling generated by the truncation and the choice of observable matter as well.
With only $C$, the more relevant $V_{0,\pm1}$ is weaker in the observed splitting than $V_{0,\pm2}$, and with only $U(1)$ the current limits the readout of the charged levels for $K\ge1.25$ even where the second-order dimerization grows faster, because the current enters at first order.
In this way the imposed symmetry controls how long a run, or a part of its spectrum, remains usable.
Imposing $O(2)$ removes the current and, for $K>2$, leaves no relevant growth at all, whereas imposing only $C$ admits $V_{0,\pm2}$ and shortens the usable range of RG steps by a factor of two to three at $K=1.5$ and $1.75$.
Because the dimerization is forbidden by one-site translation, a LoopTNR scheme preserving the original $\Tr$ together with $O(2)$ should be free of relevant perturbations for $K>1/2$, leaving only the marginal drift of $K$ and the lattice anisotropy.
Such a scheme is not easy to construct, since the original translation has no local action on the coarse-grained network (Appendix~\ref{app:lattice-sym}).
A simpler route, studied in the companion paper~\cite{Paper1b}, is to cancel the dimerization with a deformation of the initial tensor that has the same quantum numbers.

For practical calculations, the results indicate when stopping the RG after a fixed number of steps is safe.
The 24 steps at $\chi=40$ of the six-vertex benchmark in Ref.~\cite{TNRKit} lie inside the $O(2)$ weak-coupling windows for $0.75\le K\le1.75$ ($n_e=27$--$71$), and at $n=24$ the charged level deviates from its CFT value by at most $0.6\%$.
This margin is not guaranteed at other bond dimensions or with less symmetry.
At $\chi=16$ the window at $K=1$ ends at $n_e=20$, and with only $C$ imposed it ends at $n_e=23$--$26$ for $K\ge1.5$ at $\chi=40$.
Because the seed scatters strongly between neighboring $\chi$, the stopping point is better judged for each run from its fitted growth term.

On the line of fixed points the relevant and marginal directions are coupled.
Through Eq.~\eqref{eq:flow}, a growing $\gd$ drives a motion along the line that at $\chi=40$ exceeds the direct marginal error by a factor of at least six (Sec.~\ref{sec:Kn}) and accounts for most of the excess of the growth rate for $K\ge1.5$.
Unlike in analytically constructed fixed-point tensors~\cite{Cheng2026}, the marginal deformation in a numerical RG is generated by the truncation, at a rate that we have measured.
In contrast to the matrix-product-state setting, a real-space RG amplifies the perturbation step by step, so that its growth rate, phase, and amplitudes can be followed, and the lattice symmetry of the coarse-graining map itself enters the selection of the perturbation.

Our conclusions rest on growth rates extracted from finite windows in the weak-coupling regime.
The error bars measure the sensitivity to analysis choices.
We have studied a single algorithm and model, have not established convergence in $\chi$, and varied the tube width only by a factor of $\sqrt5$ (and of two in the independent runs).
Other coarse-graining schemes, such as HOTRG~\cite{Xie2012}, TNR~\cite{Evenbly2015}, or GILT~\cite{Hauru2018}, preserve different lattice symmetries, so that the selected operators may differ there, although the method of identification carries over.
Near $K=2$, where the dimerization becomes marginal, the coupled nonlinear flow of $(\gd,K)$ would have to be treated explicitly.
Two discrepancies remain unexplained: an excess of about $3\%$ in the amplitude ratio at $K=1.75$ that persists down to the cut $|X|\le0.01$ for $\chi\ge48$ (Appendix~\ref{app:k175}), and the growth of the dimerization at the BKT point, which falls short of its logarithmically corrected value by $1.5$--$5.5\%$, depending on how the background and the running rate are treated (Sec.~\ref{sec:bkt}).

The main open question is what sets the size of the seed.
From $\chi=16$ to $64$ it decreases by about four orders of magnitude for $K=1$--$1.5$ and by about three at $K=0.75$, but not monotonically, and it fluctuates by up to two orders of magnitude between neighboring $\chi$.
At $K=1.75$ it shows no systematic decrease for $\chi=24$--$48$ (Sec.~\ref{sec:chi}).
We have no quantitative model of it, nor of the seeds of $V_{0,\pm2}$ and $V_{0,\pm1}$ with only $C$ or of the $U(1)$ current.
The companion paper~\cite{Paper1b} will address where the dimerization seed is injected.
These open points do not affect the identification of the growing operators, which is based on rates, amplitude ratios, and selection rules that are consistent across all runs.

The same diagnostics can be applied to other $\mathsf{c}=1$ models.
In the Ashkin--Teller model~\cite{AshkinTeller1943,Chatelain2025}, whose internal symmetry $D_4$ does not forbid the thermal operator, criticality is protected by self-duality, so the question is whether a finite-$\chi$ RG can preserve it.
In the XY model, the low-temperature phase is a second realization of the line of fixed points, and the BKT point allows a comparison with level spectroscopy~\cite{Ueda2021}.

\begin{acknowledgments}
  This research was supported in part by the World Premier International Research Center Initiative (WPI), MEXT, Japan; JSPS KAKENHI Grant No.~23K25865; JST, Japan (CREST Grant No.~JPMJCR26XA, Moonshot R\&D Grant No.~JPMJMS256E); and IBM-UTokyo-sponsored research.
  This work was carried out as part of the Google UTokyo Quantum Partnership Research Program.
  This research in part used computational resources of Miyabi provided by Multidisciplinary Cooperative Research Program in Center for Computational Sciences, University of Tsukuba.
  The authors used AI tools (Anthropic Claude Opus~5.5 via Claude Code; OpenAI Codex/ChatGPT with GPT-6.1 Sol/Astra; Google Gemini~4 Hydrogen/Argon via Antigravity) to draft and revise the text of this manuscript from the authors' notes, to organize the literature, to write and debug the run, analysis, and plotting scripts, and to check the manuscript against the data and scripts.
  All text, claims, and references were checked and revised by the authors.
\end{acknowledgments}

\paragraph*{Data availability.}
The transfer-matrix spectra, run and analysis scripts, and figure data underlying this paper are available at \url{https://github.com/masahitoyamazaki/looptnr-c1-spectroscopy}; the repository also documents each run type of Appendix~\ref{app:numerics}. The coarse-grained tensors are available from the authors upon reasonable request.

\appendix

\section{Lattice symmetries and the height representation}
\label{app:lattice-sym}

We use the height representation of the six-vertex model~\cite{vanBeijeren1977,Baxter1982}.
Integer heights $\hat h$ are assigned to the faces of the lattice such that crossing a bond changes $\hat h$ by $+1$ ($-1$) if the arrow points to the right (left) of the direction of motion.
The ice rule ensures that $\hat h$ is single valued around each vertex.
Since every step changes $\hat h$ by $\pm1$, the parity of $\hat h$ alternates between the two sublattices of faces.
We take $\hat h$ even on one fixed sublattice, so that the arrows determine $\hat h$ up to a global shift by an even integer.
On the lattice, $e^{i\pi e\hat h}=(-1)^{e\hat h}$ is real and cannot distinguish $e$ from $-e$.
This distinction appears only at long distances, where the heights are coarse-grained into a real-valued field $\mathsf h$ proportional to the field $\phi$ of Sec.~\ref{sec:model}.
We normalize it such that the vertex operators with $m=0$ are $V_{e,0}\propto e^{i\pi e\mathsf h}$, i.e., $\mathsf h=\sqrt K\phi/\pi$.
For this field, $V_{e,0}$ and $V_{-e,0}$ are independent, and so are $\Vd$ and $\Vs$.
A lattice symmetry acts on $\hat h$ as $\hat h\to\pm\hat h+{\rm const}$.
Since coarse-graining commutes with such maps, $\mathsf h$ transforms in the same way, and the transformation of $V_{e,m}$ follows from that of $\mathsf h$.
A nonzero $m$ corresponds to a height mismatch $2m$ around the cylinder, i.e., to the $U(1)$ charge.
This charge is half the difference between the numbers of up and down arrows in a row of vertical bonds, which is $S^z$ in the XXZ language.

\emph{Charge conjugation} $C$ reverses all arrows, which maps $\hat h\to-\hat h$ and therefore $\mathsf h\to-\mathsf h$, as well as $m\to-m$, so that $(e,m)\to(-e,-m)$.
It also changes the sign of the currents.
\emph{One-site translation} $\Tr$ preserves the orientation rule, maps each face to a face of the other sublattice, and therefore maps $\hat h\to\hat h+{\rm odd}$, hence $\mathsf h\to\mathsf h+{\rm odd}$, which gives $V_{e,m}\to(-1)^eV_{e,m}$.
In the XXZ language, $(-1)^e$ is the momentum $\pi$ of the dimerization and staggered fields.
In the frame used here, in which the $XY$ coupling is ferromagnetic, $S^\pm$ carries no momentum.
The familiar momentum $\pi$ of $S^\pm$ belongs to the antiferromagnetic frame, which differs from ours by a rotation of every second spin.
The additional phase $(-1)^m$ under $\Tr$ in Ref.~\cite{Cheng2023}, whose XXZ chain has an antiferromagnetic $XY$ coupling, arises from this difference in frame.
\emph{Rotation} by $90^\circ$ is a symmetry for $a=b$ and preserves the orientation rule of the heights, so an operator of conformal spin $s$ acquires the factor $i^s$.
For a rotation about the center of a face, which preserves the face sublattices, this is the only factor.
A rotation about a vertex permutes the four faces around it cyclically, exchanges the two face sublattices, and maps $\hat h\to\hat h+{\rm odd}$.
It therefore gives an additional factor $(-1)^e$, and $\Vd$ is odd under it.
\emph{Reflection} $R_{\rm v}$ about a vertical line through a column of vertices reverses the horizontal arrows and maps $a\leftrightarrow b$ (a symmetry for $a=b$).
It also reverses the orientation rule, so that $\hat h\to-\hat h+{\rm const}$ and $\mathsf h\to-\mathsf h+{\rm const}$.
The constant is odd, because $R_{\rm v}$ maps each face to a face of the other sublattice.
The vertical arrows, and hence $m$, are unchanged, and we obtain $V_{e,m}\to(-1)^eV_{-e,m}$ (consistent with the reversal of the spin $em$) and $\Vd\to-\Vd$.
A reflection about a line through a column of faces instead preserves the face sublattices and leaves $\Vd$ invariant, while it exchanges the two sublattices of vertices.
This decides which lattice deformations of the initial tensor carry the quantum numbers of $\Vd$: vertex weights that alternate between the sublattices of vertices are invariant under $R_{\rm v}$ if they are symmetric in $a$ and $b$ (e.g., an alternating $\cv$) and odd if they are antisymmetric (an alternation of $a-b$), so only the latter has the quantum numbers of $\Vd$.

On the coarse-grained network the loss of $\Tr$ can be stated precisely.
After $n$ steps the coarse lattice vectors have length $2^{n/2}a_0$, and $\Tr$ is a translation by a fraction $2^{-n/2}$ of a coarse lattice vector.
Its only action on the coarse network is through the retained bond spaces, and the truncations do not keep track of it.
The conditions $T_A=T_B$ or $C_4$ invariance of the coarse tensors constrain translations by coarse lattice vectors and rotations, but not $\Tr$.
Imposed together, they generate rotations about the centers of coarse faces.
The action of these rotations on $\Vd$ depends on how the centers sit relative to the original lattice, which we have not analyzed.
A point-group symmetry of the coarse tensors forbids $\Vd$ only if it exchanges the two sublattices of the original faces, as for $R_{\rm v}$ or a rotation by $90^\circ$ about a vertex.
No point-group symmetry is imposed in the scheme used here.
Preserving the original $\Tr$ exactly would require carrying an explicit representation $U$ of $\Tr$ on the retained spaces through every truncation.
This enlarges the local cell, and the translation-constrained variational problem is in general harder to optimize than the unconstrained loop optimization.
We have not attempted it, and a practical $\Tr$-preserving LoopTNR remains an open problem.

\section{Second-order coefficients}
\label{app:coefficients}

On a cylinder of circumference $2\pi$, write $\mathcal H=\mathcal H_{\rm CFT}+\gd\mathcal V_{\mathrm d}$ with $\mathcal V_{\mathrm d}=\int_0^{2\pi}d\theta\,\Vd^{\rm cyl}(\theta)$.
The second-order shift of a level measured from the vacuum is
\begin{equation}
  \alpha_i=\sum_{\psi\notin\mathcal D_i}\frac{|\langle \psi|\mathcal V_{\mathrm d}|i\rangle|^2}{x_i-x_\psi}
  -\sum_{\psi\neq0}\frac{|\langle \psi|\mathcal V_{\mathrm d}|0\rangle|^2}{-x_\psi},
  \label{eq:alphasum}
\end{equation}
where $\mathcal D_i$ is the degenerate subspace of $i$.
Equivalently,
\begin{equation}
  \begin{aligned}
    \alpha_i=-\pi\!\int\! d^2z\, & |z|^{-2}z^{K/2}\bar z^{K/2}                                                       \\
                                 & \times\bigl[\langle i|\Vd(z)\Vd(1)|i\rangle-\langle0|\Vd(z)\Vd(1)|0\rangle\bigr],
  \end{aligned}
\end{equation}
which is evaluated with the Coulomb-gas integrals of Dotsenko and Fateev~\cite{Dotsenko1984,Dotsenko1985} and analytic continuation in $K$ where the integral diverges.
Consider a primary state $|e,m\rangle$ with chiral momenta $p=e\sqrt K+m/(2\sqrt K)$ and $\bar p=e\sqrt K-m/(2\sqrt K)$ [so that $x_{e,m}=\tfrac12(p^2+\bar p^2)$ and $s=em$], and normalize $\langle V_{1,0}(z)V_{-1,0}(0)\rangle=|z|^{-2K}$.
The term $V_{\pm1,0}(z)V_{\mp1,0}(1)$ of $\Vd(z)\Vd(1)$ then has the matrix element
\begin{equation}
  \begin{aligned}
     & \langle e,m|V_{\pm1,0}(z)V_{\mp1,0}(1)|e,m\rangle     \\
     & \quad=z^{\pm(eK+m/2)}\bar z^{\pm(eK-m/2)}|1-z|^{-2K},
  \end{aligned}
\end{equation}
while $V_{\pm1,0}V_{\pm1,0}$ changes $e$ by $\pm2$ and contributes only between the degenerate states $(\mp1,0)$ and $(\pm1,0)$, which are treated separately below.
The integral then reduces to
\begin{equation}
  \begin{aligned}
     & \int d^2z\,z^a\bar z^{\bar a}(1-z)^b(1-\bar z)^{\bar b}                                                     \\
     & \quad=\pi\frac{\Gamma(1+a)\Gamma(1+b)\Gamma(-1-\bar a-\bar b)}{\Gamma(-\bar a)\Gamma(-\bar b)\Gamma(2+a+b)}
  \end{aligned}
\end{equation}
($a-\bar a,\,b-\bar b\in\mathbb Z$).
For $a=\bar a$ and $b=\bar b$ this is Eq.~(B.9) of Ref.~\cite{Dotsenko1985} with $m=1$ in the notation of that reference.
Its extension to integer $a-\bar a$ and $b-\bar b$ follows from the same contour deformation, and we have checked it by numerical integration.
We need it for $a=\tfrac K2-1\pm(eK+\tfrac m2)$, $\bar a=\tfrac K2-1\pm(eK-\tfrac m2)$, and $b=\bar b=-K$.
The two signs give the same value, since the result is invariant under $e\to-e$ and, by $\Gamma(1+a)\Gamma(-a)=-\pi/\sin\pi a$, under $m\to-m$.
The primary states $(e,m)$ other than $e=\pm1$, $m=0$ therefore give
\begin{equation}
  \begin{aligned}
    \alpha_{e,m} & =-\pi\left[I_{e,m}-I_{0,0}\right],                                              \\
    I_{e,m}      & =\pi\frac{\Gamma(1-K)}{\Gamma(K)}
    \frac{\Gamma(\frac K2+eK+\frac m2)}{\Gamma(1-\frac K2+eK+\frac m2)}                            \\
                 & \quad\times\frac{\Gamma(\frac K2-eK+\frac m2)}{\Gamma(1-\frac K2-eK+\frac m2)},
  \end{aligned}
\end{equation}
and, with $\mathcal G(z)\coloneqq\Gamma(z)/\Gamma(1-z)$, the $C$-even (E) and $C$-odd (O) combinations of $e=\pm1$, $m=0$ give
\begin{equation}
  \begin{aligned}
    \alpha^{\rm E/O}_{1,0}=-\pi^2\Bigl[&\mathcal G(\tfrac{3K}{2})\mathcal G(1-K)\mathcal G(-\tfrac K2)\\
                                   &\pm\tfrac12\mathcal G(-\tfrac K2)^2\mathcal G(1+K)-\mathcal G(\tfrac K2)^2\mathcal G(1-K)\Bigr].
  \end{aligned}
\end{equation}
Descendant levels are computed with the free-boson Wick theorem in the same way.
The pole of Sec.~\ref{sec:cpt} provides an internal check, since two levels with different $\partial_Kx$, whose coefficients are computed independently, must give the same residue.
For $K\to2$ we find $(2-K)\alpha_{0,1}\to\pi^2/4$ and $(2-K)\alpha^{\rm O}_{1,0}\to-4\pi^2$, where $\alpha^{\rm O}_{1,0}$ refers to the $C$-odd combination of $e=\pm1$, as required by $\partial_Kx_{0,1}=-1/(4K^2)$ and $\partial_Kx_{1,0}=1$.

The coefficients entering Eq.~\eqref{eq:ampratio} can also be checked on the lattice, independently of the Coulomb-gas integrals.
Consider the periodic XXZ chain with bond alternation $1\pm\delta$, $\Delta_{\rm XXZ}=-\cos[\pi/(2K)]$, and length $L_{\rm ch}$.
The second-order shifts of the lowest charged level and of the current level grow as $\delta^2L_{\rm ch}^{2(2-K)}$, and the ratio of their growing parts must equal $\alpha_J/(4K\alpha_{0,1})$.
We fit the shifts as $b^{(0)}_i+b^{(1)}_iL_{\rm ch}^{2(2-K)}$ ($i=X$ for the charged level, $J$ for the current level) and compare the ratio $b^{(1)}_J/b^{(1)}_X$ with the prediction.
Exact diagonalization for $L_{\rm ch}=12$--$24$ gives $0.3606$ at $K=1$ (prediction $0.3607$), $0.3230$ at $K=1.25$ ($0.3206$), and $0.3869$ at $K=0.75$ ($0.3924$), which agree within $1.4\%$.
At $K=1.5$ the growing part ($\propto L_{\rm ch}$) is only weakly separated from the size-independent part, and $L_{\rm ch}=12$--$24$ gives $0.281$ (prediction $0.264$).
For larger systems we use density-matrix renormalization group (DMRG) calculations for $L_{\rm ch}=24$--$48$, with bond dimension 256 ($L_{\rm ch}\le40$) and 400 ($L_{\rm ch}=40$, $48$).
At $L_{\rm ch}=40$, increasing the bond dimension from 256 to 400 changes the shift of the charged level by a relative $10^{-5}$ and that of the current level by $8\times10^{-4}$.
These calculations bring the ratio to $0.268$ ($L_{\rm ch}=24$--$48$) and $0.266$ ($L_{\rm ch}=40$--$48$).
A fit that includes a $1/L_{\rm ch}$ correction gives $0.264$ ($L_{\rm ch}=12$--$24$) and $0.263$ ($L_{\rm ch}=24$--$48$), in agreement with the prediction.
For the exact diagonalization we use sparse Lanczos in the sectors $S^z=0,1$, identify the current pair by its eigenvalue $e^{\pm4\pi i/L_{\rm ch}}$ under translation by two sites, and obtain the $\delta^2$ coefficient from $\delta=2\times10^{-3}$ and $4\times10^{-3}$ by Richardson extrapolation.
The DMRG uses TeNPy~\cite{Hauschild2018,Hauschild2024} (finite matrix product states with periodic couplings, $S^z$ conserved, excited states by orthogonalization), with $\delta=5\times10^{-3}$ and $10^{-2}$.

We checked the procedure against Table~II of Ref.~\cite{Ueda2023}.
Applied to the Ising model ($\mathsf{c}=1/2$), it reproduces the first-order corrections from the dimension-four operator, $-7\pi/768$ for $\sigma$ and $7\pi/48$ for $\varepsilon$, whereas the second-order coefficients are left unevaluated there.
Our second-order coefficients agree to 12 digits with the exact free-fermion values at $K=1$ ($\alpha=4\pi^2$ for $J$, $e=\pm1$, and $m=2$; $8\pi^2/3$ for $T$; $8\pi^2$ for $J\Jb$; $4\pi^2\ln2$ for $m=1$), and they reproduce the $SU(2)$ multiplet structure at $K=1/2$.
Table~\ref{tab:alpha} lists the values used in this paper.

\begin{table}[tb]
  \caption{Second-order dimerization coefficients in the normalization of Eq.~\eqref{eq:coupling}: lowest charged level ($\alpha_{0,1}$), current ($\alpha_J$), and $C$-even/odd $e=\pm1$ states.}
  \label{tab:alpha}
  \begin{ruledtabular}
    \begin{tabular}{lcccc}
      $K$  & $\alpha_{0,1}$ & $\alpha_J$ & $\alpha^{\rm E}_{1,0}$ & $\alpha^{\rm O}_{1,0}$ \\
      \hline
      0.60 & 54.446         & 53.542     & 113.766                & 58.321                 \\
      0.75 & 40.640         & 47.840     & 83.533                 & 48.783                 \\
      1.00 & 27.364         & 39.478     & 39.478                 & 39.478                 \\
      1.25 & 20.320         & 32.578     & $-20.060$              & 30.969                 \\
      1.50 & 17.089         & 27.059     & $-141.304$             & 12.289                 \\
      1.75 & 18.847         & 22.805     & $-528.047$             & $-59.409$              \\
    \end{tabular}
  \end{ruledtabular}
\end{table}

\section{Numerical details and fit protocols}
\label{app:numerics}

Besides the main runs (Sec.~\ref{sec:method}), we use additional runs with the same code and settings but with options that are inactive by default (other values of $K$ or $\chi$, other imposed symmetries, and readouts of saved tensors).
Two further data sets come from separately scripted runs and are identified where they are used.
Finally, an independent set of runs with a development version of the code agrees with the main runs where they overlap (Appendix~\ref{app:earlier}).
We describe each of them below.

\emph{Main runs:} Julia~1.13.3, TNRKit~0.7.0 (released version) and TensorKit~0.16.5, default \texttt{LoopParameters()}, and the readout with fixed $\tau=i/2$ (eigenvalue phases stored).
We ran $O(2)$ and $U(1)$ at $K=0.75$, $1$, $1.25$, $1.5$, $1.75$, $2.5$, each for $\chi=16$, $24$, $32$, $40$, $48$, $64$.
The initial tensor is the vertex tensor of the six-vertex model in the arrow basis, constructed by TNRKit with the imposed symmetry.
Its $U(1)$ and $O(2)$ constructors differ by an exchange $b\leftrightarrow\cv$ of the weights, which we verified by exact contraction on $2\times2$ and $2\times4$ tori.
All values of $K$ refer to the physical weights $(a,b,\cv)=(1,1,\cv)$ of Sec.~\ref{sec:model}.
The runs extend to 80 RG steps, or stop earlier when the central charge has stayed below $0.1$ for three steps or the readout fails, both of which happen well after the breakdown of the weak-coupling window.
They were performed on a 64-core workstation and on Miyabi-C at JCAHPC with identical package environments.
Their scripts, environment files, and analysis are part of the data repository (see Data availability).
With 5--11 BLAS threads per run, one RG step takes about $12$~s at $\chi=40$ and $90$--$110$~s at $\chi=64$, and the readout $10$--$90$~s.
Loop optimization stops when the normalized squared error $\|\Psi-\tilde\Psi\|^2/\|\Psi\|^2$ between the loop $\Psi$ of four tensors before truncation and its truncated approximation $\tilde\Psi$ falls below $10^{-9}$ or after 20 sweeps.
This is a stopping criterion and does not guarantee convergence at every step.
The entanglement filtering uses a relative singular-value cutoff of $10^{-14}$.

The central charge $\mathsf{c}$ is obtained from the leading eigenvalue $\lambda_0$ as $\mathsf{c}=(6/\pi)\ln\lambda_0/(\operatorname{Im}\tau_n-\mathcal A/4)$~\cite{TNRKitCode}.
Here the tensors are normalized by the fourth root of the leading eigenvalue of the minimal two-row transfer matrix of $T_A$ and $T_B$, which removes the bulk free energy.
The area of the tube in units of $a_n^2$ is $\mathcal A=4$, and for the standard tube this gives $\mathsf{c}=-(12/\pi)\ln\lambda_0$.
Eigenvalues are computed with a Krylov method (Krylov dimension $\ge40$, at most 100 iterations, tolerance $10^{-12}$).
We request eight levels per sector: the neutral $C$-even and $C$-odd sectors and $|m|=1,2,3$ for $O(2)$, and $q=0,\pm1,\pm2,\pm3$ for $U(1)$.

\emph{Additional runs} use the same environment, code, and readout, with options of the run script that are inactive by default.
(a) Other parameters: $O(2)$ at $K=0.5$ and $0.6$ ($\chi=24$--$64$), at $K=2.1$ and $2.25$ ($\chi=40$ and $64$), and at $K=1.25\pm0.001$ and $2.5\pm0.001$ ($\chi=40$), and $O(2)$ with $\chi=96$ at $K=1$ and $1.5$.
(b) Tensors with only $C$ ($\mathbb Z_2$-graded in the eigenbasis of arrow reversal, 16 levels per sector).
(c) Readouts of saved tensors: the saved $U(1)$ tensors at $K=2.5$ with 24 neutral levels, and the saved $O(2)$ tensors with the wide tube (see below).
At $K=2.1$, $2.25$, and $2.5$ the main and additional runs show no growth over 80 steps, consistent with the independent runs (Appendix~\ref{app:earlier}).
In these runs $|X|$ stays below $3.2\times10^{-3}$ for $\chi=40$ and $7\times10^{-4}$ for $\chi=64$ and drifts linearly, by $-(3$--$4)\times10^{-5}$ and $-(0.7$--$0.9)\times10^{-5}$ per step, respectively, with $|\mathsf{c}-1|\le6\times10^{-4}$ from step 10 on.
The additional runs at $K=0.6$ ($\chi=24$--$64$) give $r_{\rm d}/2^{2-K}=0.960$--$0.974$ and a normalized amplitude ratio of $0.985$--$0.997$, consistent with the denser scan of the independent runs.
In these runs the umklapp rate $2^{2-4K}=2^{-0.4}$ is distinct from the dimension-four rates.

\emph{Separate runs.}
Two data sets come from runs with their own scripts, of which we use only the stored spectra.
These are a $C$-only scan at $\chi=16$ over 17 values of $K$ and $C$-only runs with a transverse field at $K=1.5$ and $\chi=16$ (Sec.~\ref{sec:conly}).
The independent runs are described in Appendix~\ref{app:earlier}.
The $C$-only splitting fit of Sec.~\ref{sec:conly} uses the contiguous steps $n\ge10$ with $10^{-5}<8K|\Delta x|<10^{-2}$.
The ranges quoted are over the bounds $\{10^{-5},10^{-4}\}\times\{3\times10^{-3},10^{-2},3\times10^{-2}\}$.

For the $O(2)$ growth-rate fits (Sec.~\ref{sec:o2-rate}), $\ln r_{\rm d}$ is scanned on a grid of 4000 points in $[\ln1.02,\ln4]$, with all amplitudes obtained by linear least squares for each $r_{\rm d}$.
The window variations are the starts $n_s=10$, $12$, $14$ combined with the cuts $|X|\le0.05$, $0.10$.

For the $U(1)$ fits, the two-level window is the contiguous range of steps from $n=15$, closed at the first failure, in which both charged levels are positive, finite, and within $10\%$ of $1/(4K)$ and $|\mathsf{c}-1|<0.2$; it is fixed independently of the fitted rate, and decaying solutions are allowed in the fit.
We optimize $(\ln r_J,\omega)$ by the Nelder--Mead method, starting from the five best points of a $9\times35$ initial grid over $-1.5\le\ln r_J\le1.5$, $0.01\le\omega\le\pi-0.01$.
The ranges quoted are over windows of $8$, $10$, and $12$ points with the end shifted by $0$, $1$, and $2$ steps.
All fits, figures, and tables are reproduced from the stored spectra by the analysis scripts.

\emph{Rotation per step.}
The rotation angle of $135^\circ$ per step is fixed by how the legs of the coarse-grained tensors are assigned to the diagonals of the original lattice.
Rotations by $45^\circ$ and $135^\circ$ produce the same square lattice but differ by a relabeling of the legs by $90^\circ$, which is not a symmetry here because $C_4$ is not imposed.
Equivalently, the modular parameter of the unit cell maps as $\tau\mapsto(1+\tau)/(1-\tau)$~\cite{TNRKitCode}.

\emph{Seed and fit protocol.}
The seed $\gd^2(12)$ of Sec.~\ref{sec:cpt} depends on the fit protocol.
If the rate is fixed to its theoretical value and the window is cut at $|X|\le0.02$, the seed changes by a $K$-dependent factor, $0.7$--$1.6$ for $K\le1.5$ and $4$--$5$ at $K=1.75$.
At fixed $K$ this factor varies by less than $8\%$ between $\chi=32$ and $40$, and by up to $30\%$ when $\chi=24$ is included.
This is small compared with the change of the seed itself, so the dependence of the seed on $\chi$ stays the same (\nolinkurl{runs/seed_refit/} in the data repository).

\emph{Block bootstrap.}
The data are deterministic, but the residuals are irregular from step to step.
To measure how sensitive the fits are to this irregularity, we use a moving-block bootstrap (300 resamplings of blocks of four steps of the paired residuals of the two levels), which treats the residuals as a stationary random sequence.
For $0.75\le K\le1.5$ and $\chi\ge32$ the resulting spread is at most $0.18\%$ in $r_{\rm d}/2^{2-K}$ and $0.6\%$ in the normalized amplitude ratio of Sec.~\ref{sec:o2-amp}.
For the growth factor this spread is smaller than the range over fit windows in all cases.
For the amplitude ratio it is larger in three cases ($K=0.75$ and $1.25$ at $\chi=48$, $K=1.5$ at $\chi=32$), where the window range is below $0.5\%$.
We quote the range over fit windows as the uncertainty.

\emph{Wide tube.}
The wide tube of TNRKit has the shape $[\ell_{\rm h},\ell_{\rm c},\ell_{\rm s}]=[4/\sqrt{10},2\sqrt{10},2/\sqrt{10}]$ for height, circumference, and horizontal shift in units of $a_n$, i.e., $\tau=(\ell_{\rm s}+i\ell_{\rm h})/\ell_{\rm c}=0.1+0.2i$.
The standard tube has $[\sqrt2,2\sqrt2,0]$ and $\tau=i/2$, so the circumference of the wide tube is $\sqrt5$ times larger.
The readout coarse-grains the step-$n$ tensors once more, truncated to $\chi$, before forming the transfer matrix, so that both tubes are read from the same tensors at each $n$.
We applied this readout to the saved $O(2)$ tensors of all runs for which they are available ($K=1$ and $1.5$ at $\chi=24$--$64$; $K=0.75$, $1.25$, and $1.75$ at $\chi=40$), from $n=6$ to a few steps beyond $n_e$.
Because $\gd$ grows with the circumference as $L^{2-K}$, the second-order deviation, proportional to $\gd^2$, should be larger in the wide tube by $5^{2-K}$ at a given step.
At the last step inside both fit windows the measured factors are $7.23$ (prediction $7.48$), $4.88$--$4.98$ ($5$), $3.35$ ($3.34$), $2.27$--$2.33$ ($2.24$), and $1.58$ ($1.50$) for $K=0.75$, $1$, $1.25$, $1.5$, and $1.75$.
A fit over the whole window with the rate fixed to that of the standard tube gives the same factors within $0.3\%$.
With the rate left free, the wide tube gives $r_{\rm d}/2^{2-K}=0.986$--$1.038$.
The normalized amplitude ratio is more sensitive to the readout.
In the wide tube it is $0.963$--$1.036$ for $0.75\le K\le1.5$ and $\chi\ge32$, compared with $0.998$--$1.008$ in the standard tube, a difference well outside the window range and the bootstrap uncertainty.

To locate its origin, we repeated the readout with the truncation of the additional step relaxed to $2\chi$ (Sec.~\ref{sec:o2-amp}; $K=1$ and $1.5$, $\chi=32$ and $40$, RG tensors unchanged).
This gives the factors $4.94$ and $4.92$ at $K=1$ and $2.30$ and $2.32$ at $K=1.5$, with $n_e$ unchanged.
The normalized amplitude ratio changes from $1.001$ and $1.006$ to $1.004$ and $1.001$ at $K=1$, and from $0.963$ and $0.969$ to $0.995$ and $1.000$ at $K=1.5$.
At $K=1.5$, $r_{\rm d}/2^{2-K}$ also moves from $1.020$ and $1.025$ to $1.016$ and $1.016$, compared with $1.014$ and $1.015$ in the standard tube.

\subsection{Independent runs with a development version of the code}
\label{app:earlier}

As a check of reproducibility we use an independent set of runs, read out with Julia~1.13.2, whose environments and file hashes are recorded in the data repository.
It comprises $O(2)$ and $U(1)$ runs at $\chi=40$ generated with a development version of TNRKit.
In the LoopTNR step the development version differs from TNRKit~0.7.0 only by argument names, convenience methods, and documentation, and the coarse-graining is algorithmically identical.
Its CFT readout, however, additionally estimates the modular parameter $\tau_0$ of the tensor instead of fixing $\tau=i/2$.
The two data sets agree closely.
At $\chi=40$ and $K=0.75$, $1$, $1.25$, $1.5$, $1.75$, the growth factors of the independent runs are $r_{\rm d}=2.329$, $1.984$, $1.687$, $1.434$, $1.215$ (main runs: $2.331$, $1.986$, $1.687$, $1.435$, $1.221$), and the normalized amplitude ratios are $1.005$, $1.006$, $0.998$, $1.000$, $1.058$ (main runs: $1.005$, $1.003$, $0.998$, $1.003$, $1.059$).
The $U(1)$ current growth factors $r_J$ at $K=0.75$--$2.5$ are $1.4089$--$1.4144$ (main runs: $1.4084$--$1.4144$).

\begin{figure}[tb]
  \centering
  \includegraphics[width=\columnwidth]{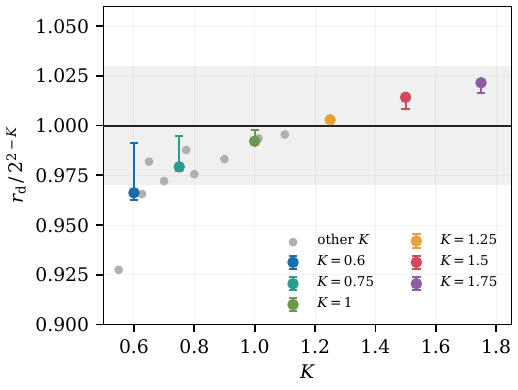}
  \caption{Independent $O(2)$ runs at $\chi=40$ (Appendix~\ref{app:earlier}): growth factor $r_{\rm d}$ of the charged-level deviation divided by $2^{2-K}$ for 15 values of $K$. Colored symbols: representative $K$ with error bars from the residual-sum-of-squares profile and window variations; gray: other $K$. The shaded band indicates $\pm3\%$.}
  \label{fig:rate-earlier}
\end{figure}

The independent $O(2)$ runs also include a denser scan of 22 values of $K$, shown in Fig.~\ref{fig:rate-earlier}.
For the 15 values $0.55\le K\le1.75$, $r_{\rm d}/2^{2-K}$ lies between $0.93$ and $1.022$, and between $0.966$ and $1.022$ for $K\ge0.6$, with the same systematic trend as in Fig.~\ref{fig:rate}.
The largest deviation, at $K=0.55$, occurs close to the BKT point $K=1/2$, where the umklapp operator becomes marginal and the fit form~\eqref{eq:fit}, which neglects its logarithmic corrections, is not expected to hold.
For the seven values $K=1.9$, $2.0$, $2.1$, $2.25$, $2.5$, $2.75$, and $3.11$, no growth rate can be identified within 70 steps.
For $K=1.9$ this is consistent with the very slow predicted growth $2^{0.1}\approx1.07$ per step.
The independent $U(1)$ runs also provide the readout with half the circumference in Sec.~\ref{sec:u1}.

\begin{figure*}[tb]
  \centering
  \includegraphics[width=0.8\textwidth]{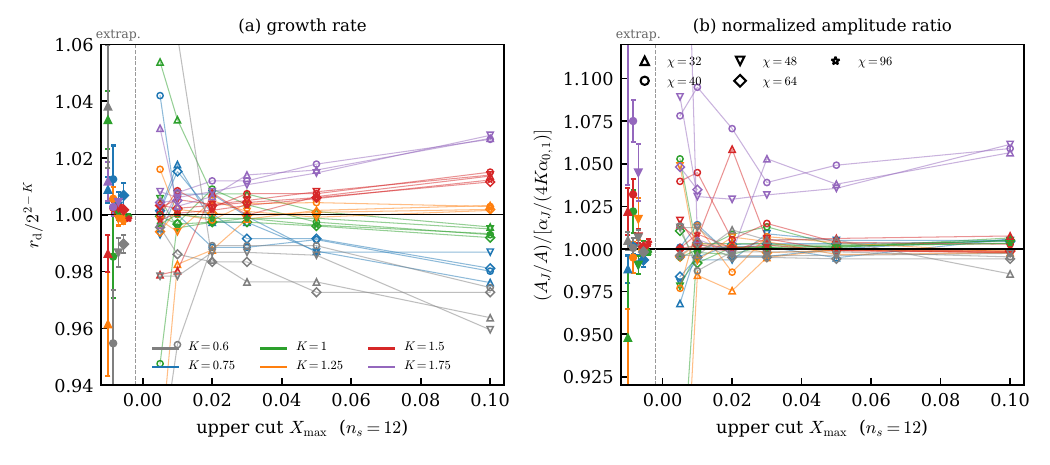}
  \caption{Dependence on the upper cut $X_{\max}$ of the fit window ($n_s=12$) of (a) the growth factor divided by $2^{2-K}$ and (b) the normalized amplitude ratio, for $O(2)$ main and additional runs with $0.6\le K\le1.75$ and $\chi=32$--$96$ (open symbols); filled symbols at the left: linear extrapolation to $X_{\max}\to0$ with indicative error bars.}
  \label{fig:window}
\end{figure*}

\subsection{Dependence on the fit window}
\label{app:window}

To see how the rate and amplitude ratio of Sec.~\ref{sec:o2} depend on the window, we refit all main and additional $O(2)$ runs with $0.6\le K\le1.75$ and $\chi\ge32$ (including $\chi=96$) with the protocol of Sec.~\ref{sec:o2-rate}.
We change only the upper cut, $X_{\max}\in\{0.1,0.05,0.03,0.02,0.01,0.005\}$.
The leading correction is of order $\gd^2\propto X$, and we extrapolate linearly in $X_{\max}$ (Fig.~\ref{fig:window}).
Since the windows are nested, the uncertainties of the intercepts are indicative.
For $0.75\le K\le1.75$ and $\chi\ge48$ the extrapolated $r_{\rm d}/2^{2-K}$ lies within $1\%$ of unity, except at $K=1.75$ and $\chi=64$, where $|X|$ reaches only $0.01$ within the 80 steps and no extrapolation is possible (e.g., $0.9989\pm0.0006$ and $0.9994\pm0.0009$ at $K=1$, $\chi=64$ and $96$; $1.0018\pm0.0004$ and $0.9986\pm0.0004$ at $K=1.5$, $\chi=64$ and $96$).
At $\chi=32$ and $40$ individual extrapolations deviate by up to $4\%$ ($0.961$ at $K=1.25$, $\chi=32$; $1.033$ at $K=1$, $\chi=32$), with no common sign.
At $K=0.6$ the deficit decreases from $3\%$ to about $1\%$, but the extrapolation is unstable.
The medians over $\chi\ge32$ are $1.008$, $0.999$, $0.999$, $1.002$, and $1.005$ for $K=0.75$, $1$, $1.25$, $1.5$, and $1.75$.
The systematic deviations of Table~\ref{tab:o2} are already gone at the cut $X_{\max}=0.02$, where at $\chi=40$ and $64$ we find $r_{\rm d}/2^{2-K}=$ $0.989$--$1.000$ at $K=0.75$, $0.997$--$1.009$ at $K=1$, $0.998$--$1.001$ at $K=1.25$, and $1.003$ at $K=1.5$.
The normalized amplitude ratio has no significant trend with $X_{\max}$ for $0.6\le K\le1.5$.
Its extrapolations are within $2\%$ of unity for $\chi\ge48$ ($1.018$ at $K=1.25$, $\chi=48$; within $1\%$ otherwise) but scatter by several percent at $\chi=32$ and $40$, where the current-level amplitude in the smallest windows is comparable to the step-to-step irregularity of that level.
At $K=1.75$ the linear extrapolation of the ratio scatters between $1.05$ and $1.12$ and is not resolved, while for $\chi\ge48$ the ratio at the cut $X_{\max}=0.01$ is about $3\%$ high (Appendix~\ref{app:k175}).
We exclude $\chi=24$, where the fits with $X_{\max}\le0.02$ are unstable for all $K$.

The analysis scripts and results are part of the data repository.

\section{Running of \texorpdfstring{$K$}{K} and the marginal directions}
\label{app:k175}

\emph{Bias of the growth rate.} To estimate the effect of the running of $K$ on the fitted rate (Sec.~\ref{sec:o2-rate}), we integrate Eq.~\eqref{eq:flow} over each RG step ($\Delta\ell=\ln\sqrt2$) at fixed $K_n$,
\begin{equation}
  \begin{aligned}
    \gd(n+1) & =2^{(2-K_n)/2}\gd(n),                                 \\
    K_{n+1}  & =K_n-\pi^2K_n^2\,\frac{2^{2-K_n}-1}{2-K_n}\,\gd^2(n),
  \end{aligned}
  \label{eq:synthetic}
\end{equation}
start from $K_{12}=K$, choose $\gd(12)$ so that the synthetic deviation matches the measured growth term at $n_e$, and fit the synthetic $X(n)$ in the same way as the data.
In the synthetic $X(n)$, the pole part of $\alpha_{0,1}$, which represents the shift of $K$ induced at fixed $K$, is replaced by the explicit running of $K$, while the regular part is kept at its value at the initial $K$.
The estimate gives $r_{\rm d}/2^{2-K}=1.005$, $1.010$, and $1.018$ for $K=1.25$, $1.5$, and $1.75$ at $\chi=40$, and a bias of at most $+0.2\%$ for $K\le1$ and $\chi\ge24$ (\nolinkurl{runs/main/analysis/synthetic_flow.py} in the data repository).

\emph{Fourth-order shift at the free-fermion point.} At $K=1$ the full dependence of the charged level on $\gd$ is known exactly from the dimerized XX chain, $X=X_g+\kappa X_g^2+\cdots$ with $X_g=-4K\alpha_{0,1}\gd^2$ and $\kappa=0.2345$, including both the running of $K$ and the fourth-order shift at fixed $K$.
A synthetic $X(n)$ built from it and fitted in the same way gives $r_{\rm d}/2=0.9924$--$0.9929$ for the windows of $\chi=32$--$64$, against the measured $0.9921$--$0.9957$, and approaches unity as the upper cut on $|X|$ is lowered, as observed (Appendix~\ref{app:window}).
Fitting the data with the rate fixed to $2^{2-K}$ and a term $\kappa X_g^2$ gives $\kappa=0.19$--$0.23$ at $K=1$ for $\chi=40$--$64$.
The same term removes the deficit at $K=0.6$ and $0.75$, with $\kappa=0.70$--$0.79$ and $0.44$--$0.51$ for $\chi=32$--$64$, independent of $\chi$, but we have not computed $\kappa$ there.

\emph{Amplitude ratio near $K=2$.} Through its pole (below Eq.~\eqref{eq:flow}), which makes up about half of $\alpha_{0,1}$ at $K=1.75$, the fixed-$K$ coefficient already contains the shift of $K$ induced at $O(\gd^2)$.
The running of $K$ is an $O(\gd^4)$ effect, enhanced near $K=2$, where the expansion parameter is the induced shift in units of $2-K$, $\varepsilon_K=\pi^2K^2\gd^2/(2-K)^2$.
Replacing the pole by the running $K(\ell)$, as in Eq.~\eqref{eq:synthetic}, gives a normalized amplitude ratio of $1.071$ at $K=1.75$ and $\chi=40$ (measured $1.059$) and $1.013$ at $K=1.5$ (measured $1.003$).
The excess at $K=1.75$ is therefore semiquantitatively consistent with a nonlinear effect of the running of $K$ and does not indicate a failure of the dimerization picture.
At the end of the standard window ($|X|=0.1$), $\varepsilon_K\simeq0.09$ at $K=1.5$ but $0.37$ at $K=1.75$, and for $|X|\le0.01$ it is $0.009$ and $0.04$.
The amplitude ratio at $K=1.5$ then stays within $1\%$ of the prediction for all cuts for $\chi\ge64$, and at the cut $|X|\le0.01$ the excess of the growth rate at $K=1.75$ shrinks from $2.7\%$ to $0.5$--$0.8\%$ for $\chi\ge40$.
The amplitude ratio at $K=1.75$, however, remains about $3\%$ high for $\chi\ge48$ at the cut $|X|\le0.01$ ($1.031$ and $1.035$ at $\chi=48$ and $64$).
At the smallest cut, $0.005$, it scatters more ($1.05$--$1.09$), and at $\chi=40$ it does not decrease with the cut.
With $d\ln[\alpha_J/(4K\alpha_{0,1})]/dK\simeq-2.8$ this corresponds to an effective $K$ lower by $\simeq0.011$.
The direct drift of Sec.~\ref{sec:Kn}, $0.003$--$0.004$ by the middle of the fit window at $\chi=40$ and smaller at larger $\chi$, accounts for at most about a third of this ($0.8$--$1.1\%$).
We do not know the origin of the rest.

\emph{Direct drift of $K$.} In the floor window, $dK_n/dn=-4.0\times10^{-4}$, $-1.5\times10^{-4}$, $-8.5\times10^{-5}$, $-4.1\times10^{-5}$, and $-1.8\times10^{-5}$ at $K=2.5$ for $\chi=24$, $32$, $40$, $48$, and $64$, and $-1.6\times10^{-4}$, $-4.7\times10^{-5}$, and $-1.5\times10^{-5}$ at $K=1.25$ for $\chi=32$, $40$, and $48$.
Accumulated to the end of the fit window at $\chi=40$, the drift gives the values of Table~\ref{tab:dK}; at $K=0.75$ the slope is positive and small, $+4.7\times10^{-6}$ per step.
The ratio of the induced to the direct shift is $0.6$--$6$ at $\chi=24$ ($K=1$--$1.75$), $1.2$--$12$ at $\chi=32$, $6$--$22$ at $\chi=40$, and $6$--$330$ at $\chi=48$ and $64$ (\nolinkurl{runs/main/analysis/drift_table.py} in the data repository).
At $K=1$, where the current and the $C$-odd combination of $e=\pm1$ are degenerate, taking the other $C$-odd level near $x=1$ (or the $C$-even level alone) for $x_e$ changes the floor-window slope by $5$--$30\%$ for $\chi=24$--$40$, the ratio at $\chi=40$ from $6.3$ to $5$--$6$, and that at $\chi=24$ to $0.5$.
With an injected shift $K\pm\delta_K$, the response $[K_n(K+\delta_K)-K_n(K-\delta_K)]/2\delta_K$ has slopes of $(+1.8\pm1.3)\times10^{-3}$ and $(-1.8\pm1.0)\times10^{-3}$ per step at $K=1.25$ and $2.5$.
The reference run at $K=2.5$ drifts by $-9.2\times10^{-5}$ per step over $n=8$--$70$ ($-8.5\times10^{-5}$ over $n\ge12$).
In the floor window the levels not used in Eq.~\eqref{eq:Kn} are predicted as $x_{0,2}=(1+\eta)/K_n$, $x_{0,3}=9(1+\eta)/(4K_n)$, and $x_J=1+\eta$.
For $\chi\ge32$ they agree with the measured ones to within $10^{-5}$--$4.4\times10^{-4}$ ($x_{0,2}$), $6\times10^{-5}$--$1.7\times10^{-3}$ ($x_{0,3}$), and $2.5\times10^{-5}$--$7.2\times10^{-4}$ ($x_J$) (rms over the window); the exception is $x_{0,3}$ at $K=0.75$ and $\chi=32$, $5.5\times10^{-3}$.
At $K=1$ the level assigned to $x_J$ is the one already used in $x_e$, so there the agreement of $x_J$ is not an independent check.
The run at $K=1.75$ and $\chi=64$ is excluded: its charged level does not reach $|X|=0.1$ within 80 steps, so that the end of its floor window is not defined.

\emph{Alternative estimators of $K$.} The injected shift of Sec.~\ref{sec:Kn} supports a marginal response but does not constrain a step-dependent bias common to nearby input values of $K$.
As a partial check of such a bias, we compare the slope of $K_n$ with those of $K_e=x_e/x_J$ and $K_m=x_J/4x_{0,1}$, which use the current level $x_J$.
These estimators are not independent, since $K_n^2=K_eK_m$.
At $K=2.5$ the slopes of $K_e$ and $K_m$ each differ from that of $K_n$ by at most $2$--$6\%$ of $|dK_n/dn|$ for $\chi=24$--$64$.
They stay negative for all nine windows tried, with start $n=12$--$20$ and end $n=60$--$80$ (\nolinkurl{external/fushio_kn_drift/} in the data repository).
At $\chi=40$ these differences are $0.9\%$ at $K=1.25$ and $7\%$ at $K=1.5$.
At $K=1.25$, $\chi=64$ and at $K=1.5$, $\chi=24$, however, the sign of the slope depends on the window.

\emph{Induced drift of $K$.} We subtract from the two levels of Eq.~\eqref{eq:Kn} the direct drift and the regular parts $(\alpha_i-\alpha_i^{\rm pole})\gd^2$ of their second-order shifts, with $\alpha_i^{\rm pole}=-\pi^2K^2\partial_Kx_i/(2-K)$ and $\gd^2(n)$ from the growth fit.
What remains is the induced shift of $K_n$ within the fit window (\nolinkurl{runs/main/analysis/induced_drift.py} in the data repository).
At its end this shift is $-0.017$ ($-0.014$) at $K=1.25$ and $-0.028$ ($-0.027$) at $K=1.5$ for $\chi=40$ ($64$), i.e., $90$--$93\%$ and $75$--$76\%$ of the fixed-$K$ shift $-\pi^2K^2\gd^2/(2-K)$ of Eq.~\eqref{eq:flow}.
This reference is not the integrated $\delta K$ of Eq.~\eqref{eq:synthetic}, in which $K$ runs.
The spectrum thus reflects the second term of the flow equation, with a deficit of $7$--$10\%$ at $K=1.25$ and $24$--$25\%$ at $K=1.5$, the larger one where $\varepsilon_K\simeq0.09$ is no longer small.
At $K=1.75$ the estimator is not reliable, because the $C$-even level in Eq.~\eqref{eq:Kn} is no longer perturbative ($4\pi|\gd|\approx2-K$).

\emph{Dimerization growth at the BKT point.} To test the growth of the dimerization at $K=1/2$ (Sec.~\ref{sec:bkt}), both the expected rate and the background must include the marginal flow.
Over the fit window $\hat g$ decreases from $0.15$ to $0.08$.
The background follows $X=\hat g/2$, since $x_{0,1}=x_t$, and we subtract $\hat g(n)/2$, with $\hat g(n)$ taken from the two-loop flow fitted to the triplet at $n=8$--$20$ and no further free parameter.
Fitting the remainder as in Sec.~\ref{sec:o2-rate}, we obtain a growth factor of $0.958$--$0.985$ times the logarithmically corrected expectation for all $\chi=32$--$64$.
If instead the factor $2^{-3\hat g(n)/4}$ is included step by step in the growth term, its prefactor is $0.945$--$0.971$ times $2^{3/2}$.
The two-loop fit to the triplet has a residual of $(2$--$4)\times10^{-4}$, and a fit of $1/\hat g$ linear in $n$ gives the same $\hat g$ within $0.001$.
Subtracting $\hat g(n)/2$ reduces the residual of the growth fit from about $10^{-3}$ to $(0.5$--$1.5)\times10^{-4}$.
With a linear background instead, the growth factor is $0.99$--$1.02$ times $2^{3/2}$ for $\chi=32$--$48$ but $1.12$ at $\chi=64$.
If the coefficient of the subtracted term is left free, the fit returns $0.52$--$0.56$ instead of $1/2$.
The window cannot be restricted to smaller $|X|$ here, since already $X(12)=\hat g/2\simeq0.074$.

\emph{Level repulsion at $K=1$.} For $\chi=32$--$64$ the splitting $S$ grows by $1.41$--$1.42$ per step in the comparison window, and floors taken over steps 8--12 instead of 8--14 change $g_S^2/g_m^2$ by less than $0.003$.
At $\chi=24$ the direct drift ($-2.7\times10^{-4}$ per step in the floor window), extrapolated from the floor (median at $n\approx10$) to the window ($n=33$--$35$), shifts $x_{0,1}$ by about $+0.0016$.
This is $10$--$26\%$ of the shifts used to define $g_m^2$ and inflates $g_m^2$ accordingly.
Removing it raises $g_S^2/g_m^2$ to $0.85$--$0.96$.

\emph{Lattice anisotropy.} The anisotropy coupled to $T,\bar T$ would appear as a period-four component of $\eta$ (Sec.~\ref{sec:cpt}).
At $K=2.5$ a fit over $n=12$--$80$ gives an amplitude of $(3$--$5)\times10^{-5}$ for $\chi\ge40$ ($1.4\times10^{-4}$ at $\chi=32$), comparable to the fit residual, so this direction is seeded only weakly.

\section{Details of the reduced-symmetry runs}
\label{app:reduced}

\subsection{\texorpdfstring{$U(1)$}{U(1)} only}
\label{app:u1checks}

\emph{Reality of the levels.} The assignment of Sec.~\ref{sec:cpt}, in which $\operatorname{Re}g_J$ splits the moduli of opposite-charge levels and $\operatorname{Im}g_J$ appears in the phases of $(\pm1,0)$, also follows from the reality of the transfer matrix.
Each $U(1)$ sector of the real transfer matrix is a real matrix, so complex-conjugate pairs of eigenvalues can occur only within one sector.
The lowest levels of opposite charge lie in different sectors and are nondegenerate within their sector, so they stay real under a small real perturbation.
The states $(\pm1,0)$, in contrast, are degenerate within the neutral sector and can form a conjugate pair.

\emph{Second-order terms.} The fields $\Vd$ and $\Vs$ do not contribute to $\Oq_q$ at second order.
For $\Vs$ alone this follows from the combined symmetry $\Tr C$, which maps $(0,q)$ to $(0,-q)$ and forces $\Oq_q=0$ to all orders.
With both $\Vd$ and $\Vs$ present, we write $\gd\Vd+g_{\mathrm s}\Vs=w_+V_{1,0}+w_-V_{-1,0}$ with $w_\pm=(\gd\mp ig_{\mathrm s})/\sqrt2$.
Conservation of $e$ allows only one $V_{1,0}$ and one $V_{-1,0}$ insertion in the diagonal second-order correction.
The coefficient of this correction is $w_+w_-=(\gd^2+g_{\mathrm s}^2)/2$, so no term proportional to $\gd g_{\mathrm s}$ appears.
The current $J$, which is $\Tr$ even and $C$ odd, breaks $\Tr C$ and splits the levels already at first order.

\emph{Tube of half the circumference.} We also read the tensors of the independent $U(1)$ runs of Appendix~\ref{app:earlier} with a tube of half the circumference.
This halves $\Oq_1$ (ratio $\rho=1.991$--$2.002$), as expected for a first-order response to a dimension-one operator.
The height and orientation of the tube also change, however, so this is not a pure test of circumference scaling.

\emph{Charge proportionality.} For the main runs of Sec.~\ref{sec:u1}, the relative residual $\epsilon_{\rm ch}$ of the fit $\Oq_q(n)\simeq q\,o_n$, $q=1,2,3$, over the fit window (the last ten points) is $24\%$, $4.2\%$, $0.31\%$, $0.26\%$, $0.13\%$, and $0.11\%$ at $\chi=40$ for $K=0.75$--$2.5$, decreasing to $0.04$--$0.8\%$ at $\chi=64$ for $K\ge1$.
Over the full two-level window it is $33\%$, $3.5\%$, $0.38\%$, $0.28\%$, $0.24\%$, and $0.09\%$ for $K=0.75$, $1$, $1.25$, $1.5$, $1.75$, and $2.5$.
The current perturbation describes the low-charge splitting well, but at small $K$ the higher charges receive additional contributions that we cannot separate from finite-tube and truncation effects.

\emph{Eigenvalue phases at $K=0.75$.} At $\chi=40$ and $64$ the two levels near $x=K$ remain real over the fit window.
This is expected when a Hermitian coupling $\delta$ between $(1,0)$ and $(-1,0)$ is present in addition to the imaginary diagonal shift $\pm i\varepsilon$, $\varepsilon=4\pi\sqrt K\operatorname{Im}g_J$ (Sec.~\ref{sec:cpt}).
It gives the pair the dimensions $K\pm\sqrt{|\delta|^2-\varepsilon^2}$, which are real as long as $|\delta|>|\varepsilon|$.
Early in the window $\delta$ comes from the slowly decaying umklapp term ($x_{\mathrm u}=4K=3$), which mixes the two states at first order, and later from the second-order dimerization, which splits the $C$-even and $C$-odd combinations of $e=\pm1$ by $(\alpha^{\rm E}_{1,0}-\alpha^{\rm O}_{1,0})(\gd^2+g_{\mathrm s}^2)$ and grows by $2^{2-K}$ per step.
At $\chi=40$ the observed half-splitting of the pair decreases from $5\times10^{-4}$ at $n=18$ to $3\times10^{-4}$ at $n=21$ and then grows to $1.27\times10^{-2}$ at $n=27$, while the predicted $\varepsilon$ grows from $0.3\times10^{-4}$ to $1.2\times10^{-3}$.
At $n=27$ the dimerization estimate $\tfrac12(\alpha^{\rm E}_{1,0}-\alpha^{\rm O}_{1,0})(\gd^2+g_{\mathrm s}^2)=1.29\times10^{-2}$, with $\gd^2+g_{\mathrm s}^2$ from the fit of $\underline{x}_1$, accounts for the observed half-splitting.
At $\chi=32$ the current coupling is more than an order of magnitude larger ($|g_J(n_e)|=1.6\times10^{-3}$ against $1.2\times10^{-4}$), and the pair becomes complex at 7 of the 10 steps of the fit window, with phases $0.93$ times the prediction~\eqref{eq:phase-pred} (correlation $0.997$).
At $\chi=48$ it is complex at $n=21$--$26$ and real at $n=27$--$29$, where the dimerization takes over.
The same mechanism explains the isolated steps at other $K$ at which the predicted phase is very small and no phase is observed.

\emph{Even part and end of the window.} We fit $X_E\coloneqq1-4K\underline{x}_1$ in the same way as the $O(2)$ charged level, from $n=12$ to the end of the two-level window.
The start of that window at $n=15$ only fixes where the search for its end begins, and the even part is already at its floor at $n=12$.
At $K=0.75$ the amplitude $\gd^2+g_{\mathrm s}^2\approx(3$--$8)\times10^{-4}$ at the end of the window is comparable to $\gd^2$ in the $O(2)$ runs at the same step.
At $K=1.25$ it is $10^{-5}$--$10^{-4}$, and for $K\ge1.5$ no growth of $\underline{x}_1$ is resolved before the current closes the window.
The fit gives $r_{\rm d}/2^{2-K}=0.991$--$0.997$ at $K=1$ ($\chi=32$--$64$) and $0.98$--$1.01$ at $K=1.25$.
At $\chi=40$ the relative shift of $\underline{x}_1$ at the last step is $9\%$ at $K=0.75$ and $8\%$ at $K=1$, compared with $0.07\%$ and $1\%$ for $\Oq_1$, so the even part ends the window for $K\le1$.
At $K=0.75$ the deficit of $r_J$ relative to $\sqrt2$ is $0.03\%$ at $\chi=48$ and $0.2\%$ at $\chi=64$.
The phase ratio of Sec.~\ref{sec:u1-phase} at $K=2.5$ is $0.996$ at $\chi=64$.

\subsection{\texorpdfstring{$C$}{C} only}
\label{app:conly}

\emph{Growth factors and windows.} At $\chi=24$ the growth factor $r_{0,2}$ of the splitting divided by $2^{1-1/(2K)}$ is $0.978$--$1.009$, except at $K=1.25$, where no window satisfies the criterion, and $K=1.75$, where $\Delta x$ changes sign within the window.
At $\chi=40$ the $C$-only runs end at $n_e=26$, $29$, $30$, $24$, $26$, and $23$ for $K=0.75$, $1$, $1.25$, $1.5$, $1.75$, and $2.5$, compared with $27$, $30$, $38$, $48$, and $71$ for the $O(2)$ runs, which show no breakdown within 80 steps at $K=2.5$ (Sec.~\ref{sec:o2-terms}).

\emph{Doublet cuts.}
The tensors are exactly $U(1)$ symmetric before the first truncation.
A truncation respecting only the $C$ grading, however, can cut through a $C$ doublet $\{|q\rangle,|{-q}\rangle\}$ of a bond (bond charges $q=0,\dots,6$ here), i.e., a $C$-even and a $C$-odd state of equal weight.
Keeping one of them mixes $q$ and $-q$, changes the charge by $2q$, and generates $V_{0,\pm2}$ but not $V_{0,\pm1}$.
In 9 of the 18 runs the first or second truncation does cut exactly through such a doublet on one to four of the eight bonds (relative splitting $10^{-11}$--$10^{-16}$).
The other nine, however, have seeds of the same size ($8K|\Delta x|\approx10^{-6}$--$10^{-4}$ at $n=8$ in both groups), so the doublet cuts do not account for the seed of $V_{0,\pm2}$.

\emph{Generation of $V_{0,\pm1}$.}
An independent analysis of a $C$-only scan at $\chi=16$ over 17 values of $K$ fits the mean $\underline{x}\coloneqq(x_{\rm even}+x_{\rm odd})/2$ and the half splitting $\Delta x$ simultaneously with the second-order dimerization, first-order $V_{0,\pm2}$, and second-order $V_{0,\pm1}$ terms.
All rates, and the ratio of the $V_{0,\pm1}$ contributions to $\Delta x$ and $\underline{x}$, are fixed by theory.
$V_{0,\pm2}$ dominates the splitting for $0.85\le K\le3$ (omitting it raises the residual by factors of $13$--$85$), and $V_{0,\pm1}$ is required near $K=2$ (factors of $17$ at $K=1.8$ and $51$ at $K=2$).
In our runs the same fit requires the $V_{0,\pm1}$ term in 6 of the 18 runs near the end of the weak-coupling window, and a second-order $V_{0,\pm2}$ term does not replace it.
In these runs its contributions to $\underline{x}$ and $\Delta x$, fitted with independent amplitudes, have a ratio of $0.36$--$0.43$ (prediction $0.371$--$0.375$).
For a deliberate injection ($K=1.5$, $\chi=16$), every bond of one orientation of the initial network is multiplied by $e^{\mu_\perp\sigma^x}$ in the arrow basis ($\mu_\perp=\pm1\times10^{-6}$, $\pm2\times10^{-6}$).
This transverse field preserves $C$ and changes $S^z$ by $\pm1$, and thus couples to $V_{0,\pm1}$ at first order.
The shift of $\underline{x}$ relative to $\mu_\perp=0$ is even in $\mu_\perp$ and increases by a factor of about four when $\mu_\perp$ is doubled (3.6 at $n=16$).
It grows by $2.9$--$4.1$ per step over $n=10$--$18$ (prediction $2^{2-1/(4K)}=3.56$) and comes with a shift of $\Delta x$ in the ratio $0.34$--$0.40$ where $|\delta \underline{x}|>10^{-2}$ (prediction $0.374$).
Extrapolated back to $n=5$, the $V_{0,\pm1}$ coupling in the six runs is $10^{-9}$--$3\times10^{-7}$, comparable to or below $g_{0,2}$.
It becomes visible only near the end of the window because it grows faster, by $2^{1+1/(4K)}\simeq2.1$--$2.5$ per step relative to the $V_{0,\pm2}$ splitting.

\clearpage
\bibliography{main_en}

\end{document}